\documentclass[11pt,a4paper]{article}

\usepackage{amsmath}
\usepackage{amssymb}
\usepackage{appendix}
\usepackage{multirow}
\usepackage{geometry}
\usepackage{float}
\usepackage{cite}

\usepackage{graphicx,axodraw}

\def\mo{\stackrel{\circ}{M}}

\newcommand{\be}{\begin{equation}}
\newcommand{\en}{\end{equation}}
\newcommand{\bea}{\begin{eqnarray}}
\newcommand{\ena}{\end{eqnarray}}

\newcommand{\mpi}{M_\pi}

\newcommand{\mpid} {M^2_\pi}
\newcommand{\mpiq} {M^4_\pi}
\newcommand{\fpid} {F^2_\pi}

\def\kmpi2{\stackrel{\circ}{M_{\pi}^2}}
\def\kmk2{\stackrel{\circ}{M_{K}^2}}
\def\kmeta2{\stackrel{\circ}{M_{\eta}^2}}

\def\kmp2{\stackrel{\circ}{M_{P}^2}}

\def\klmpip{\stackrel{\circ}{\stackrel{\sim}{M_{P}^2}}}
\def\klmpi2{\stackrel{\circ}{\stackrel{\sim}{M_{\pi}^2}}}
\def\klmk2{\stackrel{\circ}{\stackrel{\sim}{M_{K}^2}}}
\def\klmeta2{\stackrel{\circ}{\stackrel{\sim}{M_{\eta}^2}}}

\def\lmpi2{{\stackrel{\sim}{M_{\pi}^2}}}
\def\lmk2{{\stackrel{\sim}{M_{K}^2}}}
\def\lmeta2{{\stackrel{\sim}{M_{\eta}^2}}}

\def\mmu2{\mu^2}

\def\mpi2{M_{\pi}^{2}}
\def\mkd{M_{K}^{2}}
\def\metad{M_{\eta}^{2}}

\def\fpi2{F_{\pi}^{2}}
\def\fk2{F_{K}^{2}}
\def\feta2{F_{\eta}^{2}}

\def\lnmpikr{\ln{\frac{\kmpi2}{\mu^2}}}
\def\lnmkkr{\ln{\frac{\kmk2}{\mu^2}}}
\def\lnmetakr{\ln{\frac{\kmeta2}{\mu^2}}}

\def\lnmpi{\ln{\frac{\mpid}{\mu^2}}}
\def\lnmk{\ln{\frac{\mkd}{\mu^2}}}
\def\lnmeta{\ln{\frac{\metad}{\mu^2}}}

\def\lnmpikrl{\ln{\frac{\klmpi2}{\mu^2}}}
\def\lnmkkrl{\ln{\frac{\klmk2}{\mu^2}}}
\def\lnmetakrl{\ln{\frac{\klmeta2}{\mu^2}}}

\def\lnmpil{\ln{\frac{\lmpi2}{\mu^2}}}
\def\lnmkl{\ln{\frac{\lmk2}{\mu^2}}}
\def\lnmetal{\ln{\frac{\lmeta2}{\mu^2}}}

\begin{document}

\begin{flushright}
LPT-ORSAY/12-98
\end{flushright}

\vspace{0.3cm}

\begin{center}
{\bf {\Large 
Determining the chiral condensate from the distribution\\ of the winding number
beyond topological susceptibility}}

\vspace{0.5cm}

V\'eronique Bernard$^{a,b}$, S\'ebastien Descotes-Genon$^{b,c}$ and Guillaume Toucas$^{b,c}$

\vspace{0.2cm}

\emph{$^a$ Univ. Paris-Sud, Institut de Physique Nucl\'eaire, UMR8608,
91405 Orsay Cedex, France}\\
\emph{$^b$ CNRS, 91405 Orsay, France}\\
\emph{$^c$ Univ. Paris-Sud, Laboratoire de Physique 
Th\'eorique, UMR 8627,
91405 Orsay Cedex, France}
\vspace{3\baselineskip}

\vspace*{0.5cm}
\textbf{Abstract}\\
\vspace{1\baselineskip}
\parbox{0.9\textwidth}
{The first two non-trivial moments of the distribution of the topological charge (or gluonic winding number), i.e., the topological susceptibility and the fourth cumulant, can be computed in lattice QCD simulations and exploited to constrain the pattern of chiral symmetry breaking. We compute these two topological observables at next-to-leading order in three-flavour Chiral Perturbation Theory, and we discuss the role played by 
the $\eta$ propagation in these expressions. For hierarchies of light-quark masses close to the physical situation, we show that the fourth cumulant has a much better sensitivity than the topological susceptibility to the three-flavour quark condensate, and thus constitutes a relevant tool to determine the pattern of chiral symmetry breaking in the limit of three massless flavours. We provide the complete formulae for  the two topological observables in the isospin limit, and predict their values in the particular setting of the recent analysis of the RBC/UKQCD 
collaboration. We show that a combination of the topological susceptibility and the fourth cumulant is able to pin down the three
flavour condensate in a particularly clean way in the case of three degenerate quarks.}
\end{center}

\clearpage


\newpage
\setcounter{page}{1}

\section{Introduction}

An accurate description of low-energy QCD remains a very challenging task for the theorists. One of its prominent features, the spontaneous breakdown of chiral symmetry leading to Goldstone bosons identified with the pseudoscalar $\pi,K,\eta$ mesons, 
can be described through an effective theory, Chiral Perturbation Theory ($\chi$PT)~\cite{Gasser:1983yg,Gasser:1984gg}. However, this tool based on the symmetries of the 
underlying theory does not allow one to determine the precise pattern of chiral symmetry breaking. 
A very helpful complement to the effective field theory approach comes from numerical simulations of Quantum Chromodynamics on the lattice, as they provide a way to determine the low-energy constants (LECs) encoding the pattern of chiral symmetry breaking through the momentum and 
quark-mass dependence of various observables related to the dynamics of pseudoscalar mesons~\cite{Colangelo:2010et} (and conversely, chiral expansions can be exploited by lattice collaborations to extrapolate their data down to physical values of the light quark masses).

Another interesting way to determine non-perturbative features of the  QCD vacuum consists in exploiting the connection between the fluctuation 
of the winding number, or topological charge, defined from the gluonic strength tensor $G$ as
\begin{equation}
Q=\int d^4x\ \omega(x)\,,\qquad \omega=\frac{1}{32\pi^2}\epsilon_{\mu\nu\rho\sigma} {\rm Tr}[G_{\mu\nu} G_{\rho\sigma}]\,,
\end{equation}
and the breaking of the anomalous axial rotation $U_A(1)$, related to the dynamics of the $\eta$ and $\eta'$ mesons~\cite{Witten:1979vv,Veneziano:1979ec,Leutwyler:1992yt,Kaiser:2000gs}. Indeed, the fluctuations of the topological charge correspond to long-distance dynamics encoded in $\chi$PT and can be monitored during the Monte Carlo evaluation of the path integral in lattice simulations. The study of topological observables is thus able to provide additional information on the 
pattern of chiral symmetry breaking (see ref.~\cite{Vicari:2008jw} and references therein). 

In a previous article~\cite{Bernard:2012fw}, we have discussed the potentiality of the variance of the topological charge, also called topological susceptibility and measured in lattice simulations, to determine the pattern of chiral symmetry breaking. We have used data available from the RBC and UKQCD collaborations (in combination with information on the  spectrum of light pseudoscalar mesons) in order to constrain this pattern. We have shown that for simulations close to the physical point, the topological susceptibility only probes the quark condensate in the chiral limit of two massless flavours ($m_u,m_d\to 0$, $m_s$ kept at its physical value), but not  the condensate in the three-flavour chiral limit ($m_u,m_d,m_s\to 0$). Indeed, in ref.~\cite{Bernard:2012fw}, the pattern of three-flavour chiral symmetry breaking was mainly constrained by the spectrum of light pseudoscalar mesons, while the topological susceptibility played only a minor part in the analysis. The two-flavour condensate has been measured through $\pi\pi$ rescattering in various low-energy processes, such as $K_{\ell 4}$~\cite{Colangelo:2001sp,DescotesGenon:2001tn,Batley:2007zz,Colangelo:2008sm,Batley:2010zza,DescotesGenon:2012gv,Batley:2012rf} and $K\to 3\pi$ decays~\cite{Batley:2005ax,Cabibbo:2004gq,Cabibbo:2005ez,Colangelo:2006va,Gasser:2011ju}. But this condensate can be significantly higher than the three-flavour quark condensate in the presence of large contributions from sea-quark $s\bar{s}$ pairs to the chiral structure of QCD vacuum~\cite{DescotesGenon:1999uh}, a scenario supported by several phenomenological analyses
($\pi K$ scattering~\cite{Buettiker:2003pp,DescotesGenon:2007ta}, scalar form factors~\cite{Moussallam:1999aq,Moussallam:2000zf,DescotesGenon:2000di,DescotesGenon:2000ct}, lattice data~\cite{Bernard:2010ex,Bernard:2012fw}).

In the present work, we will investigate a slightly more sophisticated topological observable, namely the fourth cumulant $c_4$ of the gluonic winding number. This quantity can be measured on the lattice~\cite{DelDebbio:2002xa,D'Elia:2003gr,Giusti:2007tu,Vicari:2008jw,Chiu:2011dz,Chiu:2008jq}, and 
we will see that it does not suffer from the same shortcomings as the topological susceptibility to determine the three-flavour chiral condensate for simulations close to the physical point.
In a similar way to the topological susceptibility~\cite{Mao:2009sy,Bernardoni:2010nf,Bernard:2012fw}, the fourth cumulant $c_4$ can be computed at NLO in  $\chi$PT, and this will be the focus of the present article.
We will compute the fourth cumulant $c_4$ in two different ways.
In Sec.~\ref{sec:effpot}, we will determine it by considering the effective potential  for constant sources and obtaining its dependence on the vacuum angle $\theta$, similarly to Ref.~\cite{Mao:2009sy}.
In Sec.~\ref{sec:diags}, we will express $c_4$ as the 4-point correlator of the winding number density $\omega$ at vanishing incoming momenta, and analyse the correlator at one loop in $\chi$PT using Feynman diagrams, so that we can single out the role of the $\eta$ propagation.
In Sec.~\ref{sec:chicond}, we analyse briefly the potentiality of the fourth cumulant to extract the pattern of three-flavour chiral symmetry breaking from lattice simulations close to the physical point, with a much greater sensitivity than the topological susceptibility. We will also focus on the expansion in two-flavour $\chi$PT, before drawing a few conclusions. Two appendices describe details of the computation using Feynman diagrams, as well as the reexpression of NLO low-energy constants in terms of physical quantities.

\section{Derivation through the effective potential}\label{sec:effpot}

\subsection{Distribution of the topological charge}

Let us first define these two quantities related to topological aspects of QCD~\cite{Leutwyler:1992yt,Vicari:2008jw,Aoki:2009mx}. The partition function of QCD in a $\theta$-vacuum with a given quark-mass matrix $M$ is defined as
\begin{equation}
Z(M,\theta)=\sum_Q e^{-iQ\theta} Z_Q(M)\,,
\end{equation}
where $Z_Q$ is the partition function for a fixed topological sector, with a fixed winding number (or topological charge) $Q$, which is an integer in presence of fermion fields in the fundamental representation.
If we work in a finite volume $V$ with an Euclidean metric, we can define the vacuum energy density as
\begin{equation}
\epsilon_{vac}(M,\theta)=-\frac{1}{V}\log Z(M,\theta)\,,
\end{equation}
and its second and fourth cumulants
\begin{equation} \label{eq:c4def}
\chi=\left.\frac{\delta^2\epsilon_{vac}(M,\theta)}{\delta \theta^2}\right|_{\theta=0}\,,\qquad
c_4=\left.\frac{\delta^4\epsilon_{vac}(M,\theta)}{\delta \theta^4}\right|_{\theta=0}\,.
\end{equation}
Both quantities can be interpreted as quantities describing the distribution
of the topological charge $Q$
\begin{equation}
\chi=\frac{\langle Q^2 \rangle_{\theta=0}}{V} \qquad c_4=-\frac{[\langle Q^4\rangle-3\langle Q^2\rangle^2]_{\theta=0}}{V}\,.
\end{equation}
$\chi$, related to  the variance of the distribution, is the topological susceptibility
and has been studied in detail in Refs.~\cite{Leutwyler:1992yt,Mao:2009sy,Bernardoni:2010nf,Bernard:2012fw}.
 $c_4$ measures the kurtosis of the distribution, i.e., its more or less peaked nature (a Gaussian distribution has a vanishing kurtosis). Both quantities can be defined non-ambiguously in terms of Green functions of scalar and pseudoscalar densities at vanishing momentum, so that they are renormalisation-group invariant~\cite{Luscher:2004fu}.

For large volume and small quark masses, the partition function is dominated by the lightest states of the theory, i.e., the pseudoscalar Goldstone bosons, and can thus be analysed in terms of Chiral Perturbation Theory. As we deal only with constant source terms (quark masses and vacuum angle), we can focus on modes with vanishing momentum, i.e. the matrix $U$ collecting Goldstone bosons become independent of $x$
\begin{equation}\label{eq:effpot}
Z(M,\theta)= \int [dU] \exp[-V \mathcal{L}(U,M,\theta)]=\exp[-V \mathcal{W}(M,\theta)]\,,
\end{equation}
where $\mathcal{L}$ is the $\chi$PT Lagrangian (which can be considered for $U$ constant) and $\mathcal{W}$ is the (Euclidean) effective potential. As shown in Ref.~\cite{Gasser:1987zq}, the effective theory of QCD in a finite volume with periodic boundary conditions amounts to the same Lagrangian as in the infinite volume. The properties of this partition function have been extensively studied in Refs.~\cite{Leutwyler:1992yt,DescotesGenon:1999zj}, and in particular the distribution of the winding number $Q$ according to the leading-order (LO) chiral Lagrangian.

\subsection{Structure of the one-loop generating functional}\label{sec:genfunc}

We can follow the arguments of Ref.~\cite{Mao:2009sy} to derive the expression of $c_4$ in an elegant way using the effective potential of the theory Eq.~(\ref{eq:effpot}). The latter can be obtained from the founding work of $\chi$PT, ref~\cite{Gasser:1984gg}, where one can find the one-loop generating functional $Z$ up to $O(\Phi^4)$ in the Minkowskian metric, with the following counting
of the sources
$a_\mu\sim p=O(\Phi)$ and $v_\mu\sim s-M=O(\Phi^2)$,
but in the absence of a source term for the vacuum angle. This is however sufficient for our purposes, as a constant
source term $\theta$ can be introduced in the effective potential via an anomalous $U_A(1)$ rotation which leaves invariant the generating functional
\begin{equation}
Z(s+ip=s_0+ip_0,\theta=\theta_0)=Z((s_0+ip_0)e^{i\theta_0/N},0)\,,
\end{equation}
as discussed in detail in ref.~\cite{Gasser:1984gg}. The generating functional has been  extended later for an arbitrary number of light flavours $N$ in refs.~\cite{Bijnens:1999hw,Bijnens:2009qm},
and will be used here  to determine the contribution  of order $O(\theta^4)$ in the effective potential. We set the sources
\begin{equation}\label{eq:sources}
v_\mu=a_\mu=0, (s+ip)=Me^{i\theta/N}\,, \qquad M={\rm diag}(m_1, \ldots m_{N})\,.
\end{equation}
It is easy to determine the classical solution to the LO equation of motion under the form
\begin{equation}
\bar{U}=(e^{i\alpha_1}\ldots e^{i\alpha_N})\,,\qquad \sum_j \alpha_j=0\,,
\end{equation}
which should minimise the LO Lagrangian
\begin{equation}\label{eq:LOmin}
{\mathcal L}_1=F_{N}^2B_{N} \sum_j m_j\cos(\phi_j)\,,
\end{equation}
with $\phi_j=\theta/N-\alpha_j$ with the condition $\sum_j \phi_j = \theta$ to be fulfilled~\footnote{For simplicity, we have chosen to apply the same phase $e^{i\theta/N}$ on each of the quark masses in Eq.~(\ref{eq:sources}), but we could have chosen any redefinition of the form
\begin{equation}
v_\mu=a_\mu=0, (s+ip)={\rm diag}(m_1 e^{i w_1 \theta}, \ldots m_{N} e^{i w_N \theta})\,, \qquad \sum_i w_i=1\,.
\end{equation}
The rest of the demonstration is unchanged, provided that one adapts the definition of
$\phi_i= w_i \theta - \alpha_j$. In particular, one notices that the effective potential is only defined in terms of $\phi_i$ (and not $\theta$ and $\alpha_i$ separately), so that our results for the topological susceptibility and the fourth cumulant are independent of the choice of weights $w_i$, as expected.}. This Lagrangian involves
the pseudoscalar decay constant $F_{N}$ and
the chiral condensate $\Sigma(N)=-F^2_{N}B_{N}$ in the chiral limit of $N$ massless flavours~\footnote{$F_{N}$ and $B_{N}$ would be denoted $F_0$ and $B_0$ in the case of $N=3$ $\chi$PT~\cite{Gasser:1984gg}, and $F$ and $B$ in the case of $N=2$ $\chi$PT respectively~\cite{Gasser:1983yg}.}.

Using a Lagrange parameter, we can determine the minimum as an expansion in powers of $\theta$~\cite{Mao:2009sy}
\begin{equation}\label{eq:phi}
\phi_i=\frac{\bar{m}}{m_i}\theta+\left[\left(\frac{\bar{m}}{m_i}\right)^3-\left(\frac{\bar{m}}{m_i}\right)\frac{\bar{m}^3}{\bar{m}^{[3]}} \right]\frac{\theta^3}{6}+O(\theta^5)\,,
\end{equation}
where we have introduced the notation for the various sums and harmonic means
\begin{equation}
\bar{m}^{[k]}\equiv \frac{1}{\sum_j \frac{1}{m_j^k}}\,,\qquad  s^{[k]}\equiv \sum_j m_j^k  \,, \qquad \bar{m}\equiv\bar{m}^{[1]}\,, \qquad s\equiv s^{[1]}\,.
\end{equation}

We can then plug this expression into the one-loop generating functional $Z$ evaluated at the point Eq.~(\ref{eq:sources}). The partition function gets contributions from two different terms $Z_t$ and $Z_u$, collecting tadpoles and unitarity contributions respectively~\cite{Gasser:1984gg,Bijnens:2009qm} (there is no contribution from the anomalous part $Z_A$)
\begin{eqnarray}
Z_t/V&\to&\sum_P \left[\frac{N}{2(N^2-1)}F_{N}^2 - \frac{\mo_P^2}{32\pi^2}\log\frac{\mo_P^2}{\mu^2}\right]\sigma_{PP}^\chi
+16 B_{N}^2L_{6;N}^r(\mu)\left(\sum_jm_j\cos\phi_j\right)^2
\\  &&-16 B_{N}^2L_{7;N}^r(\mu)\left(\sum_jm_j\sin\phi_j\right)^2
  +8B_{N}^2 L_{8;N}^r(\mu)\sum_jm_j^2\cos 2\phi_j \,,\nonumber \\
Z_u/V&\to&\frac{1}{4}\sum_{P,Q} \int dx\ J_{PQ}^{r}(x)\sigma_{PQ}^\chi \sigma_{QP}^\chi\,,
\end{eqnarray}
with 
\begin{eqnarray}
\sigma^\chi_{PQ}&=&\frac{B_N}{4}\langle \{T_P,T_Q^\dag\} (uMue^{-i\theta/N}+u^\dag M u^\dag e^{i\theta/N})\rangle-\delta_{PQ} \mo_P^2
=\frac{B_N}{2}\langle \{T_P,T_Q^\dag\}A\rangle\,,\\
u&=&(e^{i\alpha_1/2}\ldots e^{i\alpha_n/2})\,,\qquad A=(m_1(\cos \phi_1-1),\ldots m_N (\cos\phi_{N}-1))\,.
\end{eqnarray}
The diagonalisation of the mass term in the LO equation of motion defines 
$\mo_P^2$, the leading-order contribution to the masses of the $N^2-1$ pseudoscalar mesons~\footnote{For $N=3$ $\chi$PT, one has
\begin{equation}
\mo_\pi^2=2mB_0\,,\qquad \mo_K^2=(m+m_s)B_0\,,\qquad \mo_\eta^2=\frac{2}{3}(2m_s+m)B_0\,.
\end{equation}},
and $T_P$, the linear combinations of the generators of the $SU(N)$ flavour group (corresponding to Gell-Mann matrices for $N=3$).
$J^r$ corresponds to the one-loop scalar integral with two mesons $P$ and $Q$,
and $L_6,L_7,L_8$ denote Low-Energy Constants (LECs) renormalised at the scale $\mu$, entering 
 the NLO chiral Lagrangian ${\mathcal L}_2$ expressed in the basis of operators detailed in Ref.~\cite{Gasser:1984gg} for $N=3$ $\chi$PT~(the same basis holds for an arbitrary number of flavours, up to the introduction of an additional LEC $L_0$ that does not enter the present computation~\cite{Bijnens:1999hw,Bijnens:2009qm}).
Using Eq.~(\ref{eq:phi}), we have the expansion of the trigonometric functions
\begin{eqnarray}
\sum_j m_j\cos\phi_j&=&s -\bar{m}\frac{\theta^2}{2}+\frac{\bar{m}^4}{\bar{m}^{[3]}}\frac{\theta^4}{24}+O(\theta^6)\,,\\
\sum_j m_j\sin\phi_j&=& N\bar{m}\theta-N\frac{\bar{m}^4}{\bar{m}^{[3]}}\frac{\theta^3}{6}+O(\theta^6)\,,\\
\sum_j m_j^2\cos2\phi_j&=&s^{[2]} -2 N \bar{m}^2 \theta^2+\frac{2}{3}N \frac{\bar{m}^5}{\bar{m}^{[3]}}
\theta^4+O(\theta^6)\,,
\end{eqnarray}
as well as the expansion of $\sigma^\chi$
\begin{equation}
\sigma^\chi_{PQ}=\frac{B_N}{2}
  \sum_{i} \{T_P,T_Q^\dag\}_{ii} m_i
  \left[-\left(\frac{\bar{m}}{m_i}\right)^2\frac{\theta^2}{2}
  +\left(\frac{\bar{m}}{m_i}\right)^2\left[4\frac{\bar{m}^3}{\bar{m}^{[3]}}- 3\left(\frac{\bar{m}}{m_i}\right)^2\right]\frac{\theta^4}{24}+O(\theta^6)
 \right]\,,
\end{equation}
and the scalar integral at vanishing transfer momentum
\begin{equation}\label{eq:unitlog}
\int dx\, J^r_{PQ}(x)=-2k_{PQ}=
   -\frac{1}{16\pi^2}\frac{\mo_P^2\log\frac{\mo_P^2}{\mu^2}-\mo_Q^2\log\frac{\mo_Q^2}{\mu^2}}{\mo_P^2-\mo_Q^2}\,,
\end{equation}
so that all the elements in $Z_t$ and $Z_u$ are simple functions of the quark masses and the chiral LECs.

\subsection{NLO expression of the topological susceptibility and the fourth cumulant}

We are now in a position to determine the one-loop expression of the first two cumulants of the winding number.
At the order $O(\theta^2)$, needed for the topological susceptibility $\chi$, one has the following contribution
\begin{eqnarray}
Z_t/V&\to&\frac{\theta^2}{2}\Bigg[
 -B_NF_N^2\bar{m} +B_N \sum_{P,i}\frac{\mo_P^2}{64\pi^2}\log\frac{\mo_P^2}{\mu^2}
    \frac{\bar{m}^2}{m_i}\{T_P,T_P^\dag\}_{ii} \\
&& \qquad  -32B_N^2L_{6;N}^r(\mu) s\bar{m}-32B_N^2L_{7;N}^r(\mu) N^2\bar{m}^2-32 B_N^2L_{8;N}^r(\mu)N\bar{m}^2
\Bigg]\,,\nonumber\\
Z_u/V&\to &0\,,
\end{eqnarray}
Using the summation formula
\begin{equation}
\sum_P \{T_P,T_P^\dag\}_{ij}=\frac{4(N^2-1)}{N} \delta_{ij}\,,
\end{equation}
the expression of the topological susceptibility at one loop given in Refs.~\cite{Mao:2009sy,Bernardoni:2010nf,Bernard:2012fw} is recovered in a straightforward way
\begin{equation}\label{eq:chit}
\chi=B_NF_N^2\bar{m} -B_N \sum_{P,i}\frac{\mo_P^2}{64\pi^2}\log\frac{\mo_P^2}{\mu^2}
    \frac{\bar{m}^2}{m_i}\{T_P,T_P^\dag\}_{ii} 
  +32B_N^2L_{6;N}^r(\mu) s\bar{m}+32B_N^2N[NL_{7;N}^r(\mu)+L_{8;N}^r(\mu)]\bar{m}^2\,.
\end{equation}
Let us notice that Ref.~\cite{Mao:2009sy} obtained this result by determining the classical solution Eq.~(\ref{eq:phi}) corresponding
to the minimum of the chiral Lagrangian up to next-to-leading order ${\mathcal L}_1+{\mathcal L}_2$. This is actually an unnecessary complication, since the expression of the one-loop effective potential
given in Ref.~\cite{Gasser:1984gg} is precisely designed to require the classical solution from the leading-order Lagrangian only. Indeed we recover the expression NLO of the topological susceptibility without being forced to perform the elaborate minimisation of Ref.~\cite{Mao:2009sy}.

In the isospin limit $m=m_u=m_d$ for $N=3$ flavours, one has~\cite{Mao:2009sy,Bernardoni:2010nf,Bernard:2012fw}~\footnote{The explanation of the superscript ``no pole'' is given in Ref.~\cite{Bernard:2012fw} and will become clear in the following.}
\begin{eqnarray}\label{eq:chinopole}
\chi^{\rm no\ pole}&=&\frac{B_0 F_0^2 m m_s}{m + 2 m_s}
+\frac{32 m m_s B_0^2 L_6^r(\mu)  (2m + m_s)}{m +  2 m_s} 
+\frac{96 m^2 m_s^2 B_0^2 [3L_7+L_8^r(\mu)] }{(m + 2 m_s)^2}\qquad\\ \nonumber
&&-\frac{3 B_0^2 m^2 m_s^2}{8\pi^2 (m + 2 m_s)^2 }\log\frac{\mo_\pi^2}{\mu^2}
-\frac{m m_s B_0^2  (m+m_s)^2}{8\pi^2(m + 2 m_s)^2}\log\frac{\mo_K^2}{\mu^2}\\ \nonumber
&&
-\frac{m m_s B_0^2  (2m+m_s)}{72\pi^2(m + 2 m_s)}\log\frac{\mo_\eta^2}{\mu^2}
 +\chi^{\rm no\ pole}d_\chi^{\rm no\ pole}\,,
\end{eqnarray}
where $d_\chi^{\rm no\ pole}$ is a remainder collecting higher-order (HO) contributions, which starts at $O(m_q^2)$.

At the order $O(\theta^4)$, the one-loop generating functional reads
\begin{eqnarray}
Z_t/V&\to& \frac{\theta^4}{24}
\Bigg[B_NF_N^2\frac{\bar{m}^4}{\bar{m}^{[3]}}
  - B_N\sum_{P,i}\frac{\mo_P^2}{64\pi^2}\log\frac{\mo_P^2}{\mu^2}
  \{T_P,T_P^\dag\}_{ii}\frac{\bar{m}^2}{m_i}\left[4\frac{\bar{m}^3}{\bar{m}^{[3]}}- 3\left(\frac{\bar{m}}{m_i}\right)^2\right] \\
&& \qquad + 32B_N^2L_{6;N}^r(\mu) \left(\frac{s\bar{m}^4}{\bar{m}^{[3]}}+3\bar{m}^2\right)
+128B_N^2L_{7;N}^r(\mu) N^2\frac{\bar{m}^5}{\bar{m}^{[3]}}
+128B_N^2L_{8;N}^r(\mu) N\frac{\bar{m}^5}{\bar{m}^{[3]}}\nonumber
\Bigg]\,,\\
Z_u/V &\to & -\frac{\theta^4}{24}
 \frac{3B_N^2}{4} \bar{m}^4 \sum_{PQ} k_{PQ} \left(\sum_i  \{T_P,T_Q^\dag\}_{ii} \frac{1}{m_i}\right)
   \left(\sum_j  \{T_Q,T_P^\dag\}_{jj} \frac{1}{m_j}\right)\,,
\end{eqnarray}
which yields the one-loop expression of the fourth cumulant
\begin{eqnarray}\label{eq:c4}
c_4&=&-B_NF_N^2\frac{\bar{m}^4}{\bar{m}^{[3]}} 
  + B_N\sum_{P,i}\frac{\mo_P^2}{64\pi^2}\log\frac{\mo_P^2}{\mu^2}
  \{T_P,T_P^\dag\}_{ii}\frac{\bar{m}^2}{m_i}\left[4\frac{\bar{m}^3}{\bar{m}^{[3]}}- 3\left(\frac{\bar{m}}{m_i}\right)^2\right] \\
&& - 32B_N^2L_{6;N}^r(\mu) \left(\frac{s\bar{m}^4}{\bar{m}^{[3]}}+3\bar{m}^2\right)
-128B_N^2N[NL_{7;N}^r(\mu) +L_{8;N}^r(\mu)]\frac{\bar{m}^5}{\bar{m}^{[3]}}\nonumber\\
&&+
 \frac{3B_N^2}{4} \bar{m}^4\sum_{PQ} k_{PQ} \left(\sum_i  \{T_P,T_Q^\dag\}_{ii} \frac{1}{m_i}\right)
   \left(\sum_j  \{T_Q,T_P^\dag\}_{jj} \frac{1}{m_j}\right)\nonumber\,.
   \end{eqnarray}

To our knowledge, Eq.~(\ref{eq:c4}) is the first NLO computation of the fourth cumulant in $\chi$PT. Let us add that this formula agrees with
the leading-order result presented in Ref.~\cite{Mao:2009sy}.
Eqs.~(\ref{eq:chit}) and (\ref{eq:c4}) could be used in principle to derive the expression of $\chi$ and $c_4$ for an arbitrary number of flavours $N$. However, one should emphasise that the notation and definitions of the LECs in these equations are derived from the basis used in the three-flavour case that will be our main focus. In Sec.~\ref{sec:twoflav}, we will discuss how these formulae must be rewritten 
to match the usual definition of the LECs in the case of $N=2$ $\chi$PT built around the two-flavour chiral limit $m_u=m_d=0$, involving only soft pions as dynamical degrees of freedom~\cite{Gasser:1983yg}.

In the case of three flavours, it is straightforward to check that our NLO expression for $c_4$ is indeed scale independent, even in the presence of strong isospin breaking. In the isospin limit $m=m_u=m_d$, one can find easily the expression for $c_4$
\begin{eqnarray}\label{eq:c4NLO}
&&c_4^{\rm no\ pole}=-\frac{B_0 F_0^2 m m_s (m^3+2m_s^3)}{(m+2 m_s)^4}-\frac{64 B_0^2 m m_s L_6^r(\mu) \left(m^4
   +2 m^3 m_s+6 m^2 m_s^2+8 m m_s^3+ m_s^4\right)}{(m+2 m_s)^4}\qquad\\
&&
   -\frac{384 B_0^2 m^2 m_s^2 (3 L_7+L_8^r(\mu)) \left(m^3+2
   m_s^3\right)}{(m+2 m_s)^5} +\frac{3 m^2 m_s^2 B_0^2 \left(m^3+2
   m_s^3\right) }{2 \pi ^2 (m+2 m_s)^5}\log \frac{\mo_\pi^2}{\mu^2}\nonumber\\
&&
   +\frac{mm_s B_0^2
   \left(m^3+9 m m_s^2+2 m_s^3\right) (m+m_s)^2 }{8 \pi ^2 (m+2 m_s)^5}\log
   \frac{\mo_K^2}{\mu^2}\nonumber\\
&&   +\frac{B_0^2 m m_s  
   \left(m^4+2 m^3 m_s+6 m^2 m_s^2+8 m
   m_s^3+m_s^4\right)}{36 \pi ^2 (m+2 m_s)^4} \log \frac{\mo_\eta^2}{\mu^2}+
   \frac{B_0^2 m^2 m_s^2 \left(13 m^2+22 m m_s+37
   m_s^2\right)}{24 \pi ^2 (m+2 m_s)^4}\nonumber\\
&& +c_4^{\rm no\ pole} d_{c_4}^{\rm no\ pole}\nonumber\,,
 \end{eqnarray} 
 where $d_{c_4}^{\rm no\ pole}$ is a remainder collecting HO contributions, starting at $O(m_q^2)$. 

The above formula features chiral logarithms of different origins. Some of them come from tadpole contributions ($Z_t$), whereas others stem from
  the loop function $J^r_{PQ}$ taken at vanishing momentum transfer as indicated in Eq.~(\ref{eq:unitlog}) ($Z_u$).
 At this order, one could in principle redefine the argument of (some of) the logarithms in order to have physical masses instead of leading order masses. This change would not affect the remainder of the NLO expansion, apart from a redefinition of HO remainders. Following the discussion in Refs.~\cite{DescotesGenon:2007ta,Kolesar:2008jr,Bernard:2010ex}), we will consider either
 the above prescription where all the logarithms have LO masses as their arguments, or 
  the one where we take physical masses for the unitary logarithms but LO ones for the tadpole logarithms~\footnote{This choice is made to meet two different criteria. First, we want to avoid redefinitions which are not motivated by physical arguments, which invites us to keep LO masses as the argument of the tadpole logarithms, as we have no intuition on how these logarithms will be affected by higher and higher-order contributions in the chiral expansion. Second, as illustrated in Refs.~\cite{Kolesar:2008jr,Bernard:2010ex}, the unitarity logarithms
 Eq.~(\ref{eq:unitlog}) arise in the momentum dependence of form factors and scattering amplitudes of pseudoscalar mesons and will induce unphysical divergences in the case of a vanishing quark condensate where $\mo_P^2\to 0$, supporting a reexpression of the arguments of the unitary logarithms in Eq.~(\ref{eq:unitlog}) in terms of physical masses.}. If
 we perform this separation, we obtain
\begin{eqnarray}\label{eq:c4NLO2logs}
&&c_4^{\rm no\ pole}=
-\frac{B_0 F_0^2 m m_s (m^3+2m_s^3)}{(m+2 m_s)^4}-\frac{384 B_0^2 m^2 m_s^2 [3 L_7+L_8^r(\mu)] \left(m^3+2 m_s^3\right)}{(m+2 m_s)^5}\\
&&
-\frac{64 B_0^2 L_6^r(\mu) m m_s \left(m^4+2 m^3 m_s+6 m^2 m_s^2+8 m
   m_s^3+m_s^4\right)}{(m+2 m_s)^4}\nonumber\\
&&+\frac{B_0^2 m^2 m_s^2
   \left(13 m^2+22 m m_s+37 m_s^2\right)}{24 \pi ^2 (m+2
   m_s)^4}+\log\frac{\mo_\pi^2}{\mu^2}\frac{3 B_0^2 m^2 m_s^2 \left(4 m^3-3 m
   m_s^2+2 m_s^3\right)}{8 \pi ^2 (m+2 m_s)^5}\nonumber\\
 &&
+\log\frac{\mo_K^2}{\mu^2}\frac{B_0^2   m m_s (m+m_s)^2 \left(m^3-3 m^2 m_s+3
   m m_s^2+2 m_s^3\right)}{8 \pi ^2 (m+2 m_s)^5}\nonumber\\
&&+\log\frac{\mo_\eta^2}{\mu^2}\frac{B_0^2  m m_s \left(2 m^4-8 m^3 m_s+13 m m_s^3+2  m_s^4\right)}{72 \pi ^2 (m+2 m_s)^4}\nonumber\\
&&+\log\frac{M_\pi^2}{\mu^2}\frac{9 B_0^2
   m^2 m_s^4}{8 \pi ^2 (m+2 m_s)^4}
+\log\frac{M_K^2}{\mu^2}\frac{3 B_0^2  m^2 m_s^2 (m+m_s)^2}{8 \pi ^2 (m+2 m_s)^4}
\nonumber\\
&&
+\log\frac{M_\eta^2}{\mu^2}\frac{B_0^2 m^2 m_s^2 (2 m+m_s)^2}{24 \pi ^2 (m+2 m_s)^4}+c_4^{\rm no\ pole} {d_{c_4}}^{\rm no\ pole}\nonumber\,.
\end{eqnarray}
It is obvious that the HO remainder $d_{c_4}^{\rm no\ pole}$ has absorbed the redefinition of the argument of the logarithms.
In the following section, we will show the computation for $c_4$ using Feynman diagrams and performing this distinction, keeping in mind that we can always revert to the prescription in Eq.~(\ref{eq:c4NLO}) by setting LO masses in the argument of all the logarithms. For both prescriptions Eqs.~(\ref{eq:c4NLO}) and (\ref{eq:c4NLO2logs}), we see that $c_4$ vanishes in the limit where $m=0$ or $m_s=0$. This is expected since the effective potential becomes independent of $\theta$ in the limit where at least one of the quark masses vanishes, as the vacuum angle can be rotated away through an anomalous $U_A(1)$ rotation (see Refs.~\cite{Mao:2009sy,Bernard:2012fw} and references therein for a more detailed discussion).

\section{Analysis in terms of Feynman diagrams}\label{sec:diags}

\subsection{Combinatorics}\label{sec:combinatorics}

\begin{figure}[t!]
\begin{center}
\includegraphics[width=12cm,angle=0]{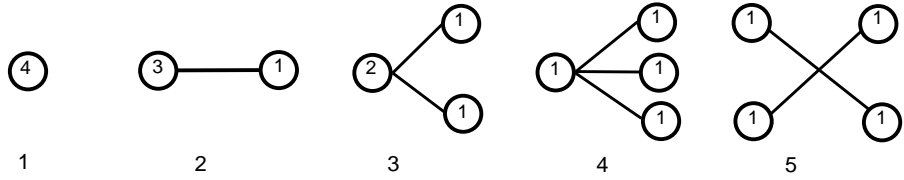}
\end{center}
\caption{\small\it Tree diagrams contributing to $c_4$. The $\theta$-induced vertices are denoted by circles containing the number of derivatives with respect to $\theta$ applied to the mass term from the LO Lagrangian in order to obtain the corresponding vertex. \label{figure:LOdiag}}
\end{figure}

It is also possible to derive the value of $c_4$ using the formalism of Feynman diagrams in $N=3$ $\chi$PT. It is obviously completely equivalent to the previous approach in terms of the one-loop generating functional,
but it allows one to separate the contributions coming from the propagation of mesons.
Computing the fourth derivative of the generating functional with respect to $\theta$, we obtain
\begin{eqnarray}\label{eq:funcderfourthderivative}
\left.\frac{\delta^4 Z}{\delta \theta^4}\right|_{\theta=0}
 &=&\int [dU] e^{iL} \Bigg[i\frac{\delta^4 L}{\delta \theta^4}
  -4\frac{\delta^3 L}{\delta \theta^3}\frac{\delta L}{\delta \theta}
  -3 \left(\frac{\delta^2 L}{\delta \theta^2}\right)^2
 -6i \left(\frac{\delta^2 L}{\delta \theta^2}\right)\left(\frac{\delta L}{\delta \theta}\right)^2
 +\left(\frac{\delta L}{\delta \theta}\right)^4\Bigg]_{\theta=0}\,,
\end{eqnarray}
with $L=\int {\mathcal L}$. At leading order, only the LO chiral Lagrangian ${\mathcal L}_1$ is needed, and more precisely, as all incoming momenta vanish, the mass term
with $Me^{i\theta/N}$, see Eq.~(\ref{eq:LOmin}). The five tree diagrams generated are indicated in Fig.~\ref{figure:LOdiag}, in the same order as the derivatives in Eq.~(\ref{eq:funcderfourthderivative}) -- the third term in Eq.~(\ref{eq:funcderfourthderivative}) cannot yield any contribution to $c_4$ at tree level, whereas the fifth term yields both diagrams 4 and 5. In Fig.~\ref{figure:LOdiag}, the number inside each circle indicates the number of derivatives with respect to $\theta$ applied to ${\mathcal L}_1$ to obtain the corresponding $\theta$-induced vertex, whereas the four-leg central vertex in diagram 5 is obtained directly from ${\mathcal L}_1$ (taking the term proportional to $B_0$, which has no derivatives). 

It is easy to compute these diagrams, with the respective contributions in the isospin limit
\begin{eqnarray}\label{eq:c4LO}
&&c_4^{\rm no\ pole}= F_0^2B_0 
 \Bigg[-
 \frac{1}{81}(2m+m_s)+\frac{8}{81}\frac{(m-m_s)^2}{m+2m_s}
  -\frac{4}{27}\frac{(m-m_s)^2}{m+2m_s}\\
&&\quad  +\frac{8}{81}\frac{(m-4m_s)(m-m_s)^3}{(m+2m_s)^3}
  -\frac{2}{81}\frac{(m-m_s)^4(m+8m_s)}{81(m+2m_s)^4}\nonumber
\Bigg]=-\frac{B_0 F_0^2 m m_s (m^3+2m_s^3)}{(m+2 m_s)^4}+O(m_q^2)\,,
\end{eqnarray}
adding up, as expected, to the LO result obtained in the previous section.
In these diagrams, the meson lines are always coupled to at least one $\theta$-induced source coming from a single derivative with respect to the vacuum angle (represented by a circle containing ``1'')
\begin{equation}
\frac{\delta {\mathcal L}_1}{\delta \theta}=\frac{F_0^2B_0}{3}\langle M\Phi\rangle + O(\Phi^3)\,,
\end{equation}
where $U=\exp(i\Phi)$. In the isospin limit, one can see that the lines involving a single meson emitted from such a $\theta$-induced vertex can correspond only to the propagation of an $\eta$ meson (there would be an additional $\pi^0$ contribution out of the isospin limit). This explains the presence of
denominators with up to four powers of $\mo_\eta^2=2(2m_s+m)/3$ in Eq.~(\ref{eq:c4LO}).

\begin{figure}[t!]
\begin{center}
\includegraphics[width=12cm,angle=0]{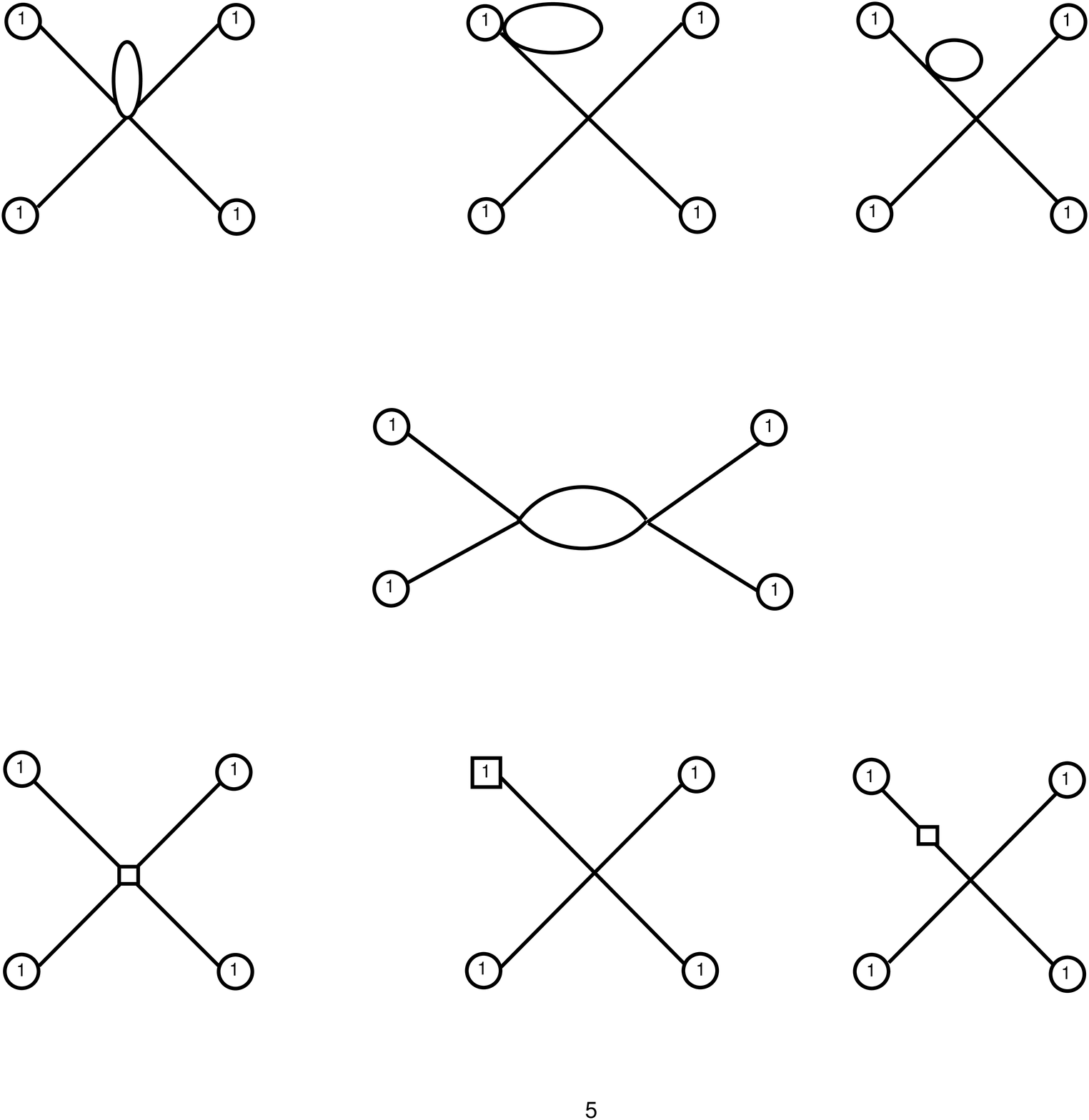}
\end{center}
\caption{\small\it  Scattering diagrams contributing to $c_4$ at one loop  and obtained by dressing the tree diagram 5, either by adding a tadpole loop (first two lines) or by replacing a LO vertex by its NLO counterpart (denoted by a square). The $\theta$-induced vertices are denoted by circles (squares respectively) containing the number of derivatives with respect to $\theta$ applied to the mass term from the LO (NLO respectively) Lagrangian in order to obtain the corresponding vertex. \label{figure:NLOdiagscatt}}
\end{figure}

\begin{figure}[t!]
\begin{center}
\includegraphics[width=12cm,angle=0]{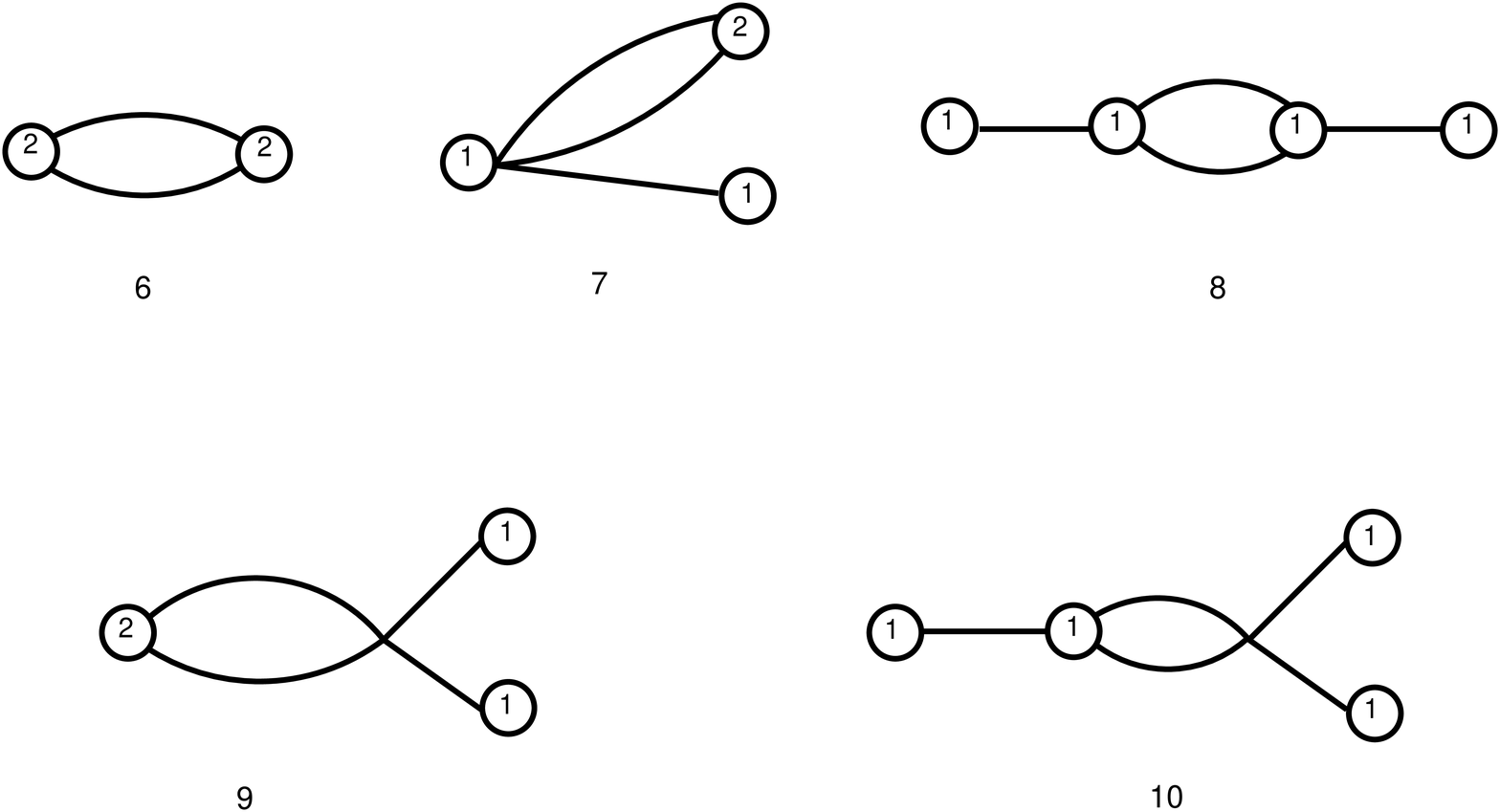}
\end{center}
\caption{\small\it  Additional diagrams contributing to $c_4$ at one loop and not derived by dressing tree diagrams. \label{figure:NLOdiagother}}
\end{figure}

Moving to next-to-leading order, we can dress the tree diagrams by \emph{a)} inserting a vertex from the NLO Lagrangian ${\mathcal L}_2$ inside one of the propagators, \emph{b)} replacing a $\theta$-induced vertex from ${\mathcal L}_1$ by its counterpart from ${\mathcal L}_2$, \emph{c)} adding a loop to one of the $\theta$-induced vertices from ${\mathcal L}_1$, \emph{d)} adding a vertex from ${\mathcal L}_1$ to build a loop either as a rescattering diagram or a tadpole on one of the propagators. These operations are illustrated in the case of
diagram 5 in Fig.~\ref{figure:NLOdiagscatt}, and the equivalent one-loop diagrams for diagrams 1-4 can be obtained straightforwardly. The number inside each circle (respectively square) indicates the number of derivatives with respect to $\theta$ applied to ${\mathcal L}_1$ (respectively ${\mathcal L}_2$) to obtain the corresponding $\theta$-induced vertex. 
 In addition to dressing the diagrams in Fig.~\ref{figure:LOdiag}, one can also take more legs out of the $\theta$-induced vertices from ${\mathcal L}_1$ and draw new diagrams, given
in Fig.~\ref{figure:NLOdiagother}. Simple parity arguments show that $\theta$-induced vertices containing an even (odd) number must have an even (odd) number of propagators attached to them.
We provide the contributions of the different diagrams in App.~\ref{app:oneloop}, and
one can check explicitly that the sum of all these diagrams yields  Eq.~(\ref{eq:c4NLO2logs}) as expected. 

\subsection{Isolating the $\eta$ propagator} 

According to the previous analysis, the powers  of $(m+2m_s)$ in the denominator of
Eq.~(\ref{eq:c4NLO2logs}) are related to the propagation of $\eta$ mesons with vanishing momentum, which are the only states coupling to the $\theta$-induced vertices in the isospin limit. Since we used NLO $\chi$PT, it is normal that the mass involved in the $\eta$-propagator is the LO value $\mo_\eta^2=2(2m_s+m)/3$. For instance, the LO expression
for $c_4$ Eq.~(\ref{eq:c4LO}) should be understood as:
\begin{eqnarray}
c_4&=&-\frac{1}{81} B_0 F_0^2 (2 m+ms)+\frac{16B_0^2F_0^2 (m-m_s)^2}{243M_\eta^2}
-\frac{16B_0^2 (m-m_s)^2(m+2m_s)}{243M_\eta^4}\\
&&+\frac{64 B_0^4F_0^2(m-4m_s)(m-m_s)^3}{2187 M_\eta^6}
-\frac{32B_0^5F_0^2(m-m_s)^4(m+8m_s)}{6561M_\eta^8}+O(m_q^2)\nonumber\,.
\end{eqnarray}
However, if we go
to higher and higher orders in Chiral Perturbation Theory, there would be tadpole and counterterm contributions to the propagator which would shift the propagator masses from LO to physical value. This is already the case with some of the contributions in diagrams 1-5, which arise when one inserts a tadpole or a NLO LEC in one propagator. These "double-propagator" contributions should actually be reabsorbed inside the mass of the $\eta$ propagator to shift its pole from $\mo_\eta^2$ to $M_\eta^2$. 

This problem is all the more acute if we do not assume that three-flavour chiral symmetry breaking is triggered by large values of the chiral condensate and the pseudoscalar decay constant, but exhibits a more complicated (and interesting) pattern. In this case, the LO contribution to the $\eta$ mass might be significantly different from its physical value, due to large contributions from NLO terms in its chiral expansion. In view of the large powers of $\mo_\eta^2$ or $M_\eta^2$ involved (up to the fourth order for diagram 5), this distinction might be quite important, and suggests that one should identify the contributions due to $\eta$ propagation in the previous computation and replace the LO $\eta$ mass by its physical value.  This point of view is in agreement with the philosophy of Resummed Chiral Perturbation Theory~\cite{DescotesGenon:1999uh,DescotesGenon:2000di,DescotesGenon:2000ct,DescotesGenon:2002yv,DescotesGenon:2003cg,DescotesGenon:2007ta,Kolesar:2008jr,Bernard:2010ex,Bernard:2012fw
}, which is built to accommodate such patterns of chiral symmetry breaking and where physically-motivated redefinitions of the chiral expansions are performed in order to limit the size of higher-order remainders.

In the present case,  it is quite straightforward for each diagram to identify the NLO contributions that should be absorbed in the shift since  the structure of the $\eta$ mass up to NLO is well known~\cite{Gasser:1984gg}. 
This yields the following expressions for the different diagrams, which should be summed together to obtain the NLO expression of $c_4$ with a physical $\eta$ propagator.

\vspace{0.2cm}

$\bullet$ Diagrams 1
\bea
&&-\frac{1}{81} B_0 F_0^2 (2 m+ms)+ \frac{ B_0^2 m^2}{216 \pi ^2} \lnmpikr
+\frac{ B_0^2
   \left(m+m_s\right){}^2}{648 \pi ^2} \lnmkkr
+\frac{ B_0^2 \left(m+2m_s\right){}^2}{5832 \pi ^2} \lnmetakr
 \\
&&
-\frac{128}{81} B_0^2  \left(m_s+2 m\right){}^2 L_6^r(\mu)
-\frac{128}{81}  B_0^2  \left(m_s+2
   m\right){}^2 L_7 -\frac{128}{81}  B_0^2  \left(2 m^2+m_s^2\right) L_8^r(\mu)\nonumber
\ena

$\bullet$ Diagrams 2
\bea
&&\frac{16B_0^2F_0^2 (m-m_s)^2}{243M_\eta^2}-\frac{ 2B_0^3  m^2 \left(m-m_s\right)}{81 \pi ^2
  M_\eta^2}  \lnmpikr
   +\frac{ 2B_0^3  \left(m-m_s\right)\left(m+m_s\right)^2}{243 \pi ^2 M_\eta^2} \lnmkkr
\\
&&
-\frac{ 2B_0^3  \left(m-m_s\right) \left(m-4m_s\right)\left(m+2m_s\right)}{2187 \pi ^2 M_\eta^2} \lnmetakr
\nonumber \\
&&
-\frac{256 B_0^3 \left(m-m_s\right)^2\left(2m+m_s\right)}{243 M_\eta^2}L_4^r(\mu)
-\frac{256 B_0^3 \left(m-m_s\right)^2\left(m+2m_s\right)}{729 M_\eta^2}L_5^r(\mu)
\nonumber\\
&&
+\frac{2560
   B_0^3 \left(m-m_s\right){}^2 \left(2m+m_s\right)}{243 M_\eta^2}
L_6^r(\mu)
+\frac{2560
   B_0^3 \left(m-m_s\right){}^2 \left(2m+m_s\right)}{243 M_\eta^2}
   L_7
\nonumber \\
&&
+\frac{2560
   B_0^3 \left(m-m_s\right){}^2 \left(m+m_s\right)}{243 M_\eta^2}
   L_8^r(\mu)
\nonumber\ena

$\bullet$ Diagrams 3
\bea
&&
-\frac{16B_0^2 (m-m_s)^2(m+2m_s)}{243M_\eta^4}+\frac{ B_0^4 m^2 \left(m-m_s\right) \left(m+m_s\right)}{27 \pi ^2 M_\eta^4} \lnmpikr
\\
&&
-\frac{ B_0^4  \left(m-m_s\right)\left(m+m_s\right)^3 }{162 \pi^2 M_\eta^4} \lnmkkr
+\frac{ B_0^4  \left(m-m_s\right) \left(m+2m_s\right) \left(m^2+m m_s-8m_s^2\right)}{729 \pi ^2 M_\eta^4} \lnmetakr
\nonumber \\
&&
+\frac{512 B_0^4 \left(m-m_s\right)^2\left(2m+m_s\right)\left(m+2m_s\right)}{243 M_\eta^4} L_4^r(\mu)
+\frac{512 B_0^4 \left(m-m_s\right)^2\left(m+2m_s\right)^2}{243 M_\eta^2} L_5^r(\mu) \nonumber\\
&&
-\frac{2040  B_0^4  \left(m-m_s\right)^2\left(m^2+m m_s+m_s^2\right)}{81 M_\eta^4}L_6^r(\mu)
-\frac{2040  B_0^4  \left(m-m_s\right)^2\left(m^2+m m_s+m_s^2\right)}{81 M_\eta^4}L_7
\nonumber\\
&&
-\frac{1024 B_0^4 \left(m-m_s\right)^2\left(m^2+m m_s+2m_s^2\right)}{81 M_\eta^4}L_8^r(\mu)
\nonumber\ena

$\bullet$ Diagrams 4
\bea
&&
\frac{64 B_0^4F_0^2(m-4m_s)(m-m_s)^3}{2187 M_\eta^6}-\frac{ 4B_0^5  m^2
   \left(4m-13 m_s\right) \left(m-m_s\right){}^2}{729 \pi ^2 M_\eta^6} \lnmpikr  
\\
&&
+\frac{2B_0^5  \left(m+m_s\right) \left(m-m_s\right)^2\left(32 m^2-99 m m_s-23 m_s^2\right)} {10935 \pi ^2 M_\eta^6} 
\lnmkkr  \nonumber\\
&&
-\frac{4B_0^5   \left(m+m_s\right) \left(m-m_s\right)^2\left(4m^2-41 m m_s+64 m_s^2\right)}{19683 \pi ^2
   M_\eta^6} \lnmetakr
\nonumber\\
&&
-\frac{1024 B_0^5 \left(m-m_s\right)^3\left(m-4m_s\right)\left(2m+m_s\right)}{729 M_\eta^6} L_4^r(\mu)\nonumber\\
&&
-\frac{1024 B_0^5 \left(m-m_s\right)^3\left(m-4m_s\right)\left(m+2m_s\right)}{2187 M_\eta^2} L_5^r(\mu) \nonumber\\
\nonumber\\
&&
+\frac{4096  B_0^5  \left(m-m_s\right)^3\left(7 m^2-11mm_s-14m_s^2\right)}{2187 M_\eta^6} L_6^r(\mu)\nonumber\\
&&
+\frac{4096  B_0^5  \left(m-m_s\right)^3\left(7 m^2-11mm_s-14m_s^2\right)}{2187 M_\eta^6} L_7
\nonumber \\
&&
+\frac{2048  B_0^5  \left(m-m_s\right)^3\left(7 m^2-9mm_s-28m_s^2\right)}{2187 M_\eta^6} L_8^r(\mu)
\nonumber\ena

$\bullet$ Diagrams 5
\bea
&&
-\frac{32B_0^5F_0^2(m-m_s)^4(m+8m_s)}{6561M_\eta^8}+
\frac{2B_0^6  m^2 \left(m-m_s\right){}^3 
\left(5m+31 m_s\right)}{2187 \pi ^2 M_\eta^8}
\lnmpikr
\\
&&
+\frac{2B_0^6  \left(m-m_s\right){}^3 \left(m+m_s\right) \left(m^2-189 m m_s+8 m_s^2\right)}{32805 \pi
   ^2 M_\eta^8} \lnmkkr
\nonumber\\
&&
+\frac{2B_0^6  \left(m-m_s\right){}^3  \left(m+2m_s\right)\left(5m^2+47 m m_s-160 m_s^2\right)}{59049 \pi ^2
   M_\eta^8} \lnmetakr
\nonumber\\
&&
+\frac{2B_0^6  m^2 \left(m-m_s\right){}^4}{729\pi ^2 M_\eta^8} \lnmpi
+\frac{2B_0^6 m^2  \left(m-m_s\right){}^4}{2187 \pi ^2 M_\eta^8} \lnmk
+\frac{2B_0^6  \left(m-m_s\right){}^4 \left(m+8m_s\right){}^2}{19683 \pi ^2 M_\eta^8} \lnmeta
\nonumber\\
&&
+\frac{2048 B_0^6 \left(m-m_s\right)^4\left(m+8m_s\right)\left(2m+m_s\right)}{6561 M_\eta^8} L_4^r(\mu)\nonumber\\
&&
+\frac{2048 B_0^6 \left(m-m_s\right)^4\left(m+8m_s\right)\left(m+2m_s\right)}{19683 M_\eta^8} L_5^r(\mu) 
\nonumber\\
&&
-\frac{1024  B_0^2  \left(m-m_s\right)^4 \left(16m^2+109 m m_s+64 m_s^2\right) }{6561 M_\eta^8} L_6^r(\mu)
\nonumber\\
&&
-\frac{4096  B_0^2  \left(m-m_s\right)^4 \left(4m^2+7 m m_s+16 m_s^2\right) }{6561 M_\eta^8} L_7
\nonumber \\
&&
-\frac{4096  B_0^2  \left(m-m_s\right)^4 \left(2m^2+9 m m_s+16 m_s^2\right) }{6561 M_\eta^8} L_8
+\frac{2B_0^6 \left(m-m_s\right){}^4 \left(37 m^2+16 m m_s+64m_s^2\right)}{19683 \pi ^2 M_\eta^8}
\nonumber\ena

$\bullet$ Diagrams 6
\bea
&&
\frac{B_0^2}{72\pi^2}  m^2  \lnmpi
+\frac{B_0^2}{216 \pi^2} \left(m+m_s\right){}^2 \lnmk
+\frac{B_0^2}{1944 \pi^2}   \left(m+2m_s\right){}^2 \lnmeta
\\
&&
+\frac{B_0^2}{1944 \pi^2} \left(37 m^2+22 m m_s+13 m_s^2\right)
\nonumber\ena

$\bullet$ Diagrams 7
\bea
&&
-\frac{ B_0^3  m^2 \left(m-m_s\right)}{27 \pi^2 M_\eta^2}  \lnmpi
-\frac{ B_0^3 m_s \left(m_s^2-m^2\right)}{81\pi^2 M_\eta^2} \lnmk\\
&&
-\frac{B_0^3  \left(m-4 m_s\right) \left(m-m_s\right)\left(m+2m_s\right)}{729 M_\eta^2\pi^2}  \lnmeta
-\frac{ B_0^2 \left(m-m_s\right){}^2 \left(17 m_s+28 m\right)}{729 \pi^2 M_\eta^2}
\nonumber\ena

$\bullet$ Diagrams 8
\bea
&&
\frac{ 2B_0^4  m^2
   \left(m-m_s\right){}^2}{81 \pi^2 M_\eta^4} \lnmpi
+\frac{ 
   2B_0^2  m_s^2 \left(m-m_s\right){}^2}{243 \pi^2 M_\eta^4} \lnmk\\
&&
+\frac{ 2B_0^2  \left(m-4 m_s\right){}^2 \left(m-m_s\right){}^2}{2187 \pi^2 M_\eta^4}  \lnmeta
+\frac{ 2B_0^2 \left(m-m_s\right){}^2 \left(28 m^2-8 m m_s+25 m_s^2\right)}
{2187 \pi^2 M_\eta^4}
\nonumber\ena

$\bullet$ Diagrams 9
\bea
&&
\frac{ B_0^4
   m^2 \left(m-m_s\right){}^2}{81 \pi^2 M_\eta^4}  \lnmpi
-\frac{ B_0^2
  m\left(m-m_s\right)^2 \left(m+m_s\right) }{243 \pi^2 M_\eta^4} \lnmk
\\
&&
+\frac{ B_0^4 \left(m+8m_s\right)\left(m-m_s\right){}^2\left(m+2m_s\right)}{2187  \pi^2 M_\eta^4} \lnmeta
-\frac{ B_0^4
  \left(m-m_s\right)^2 \left(m+m_s\right)^2}{162 \pi^2 M_\eta^4} \lnmkkr
\nonumber\\
&&
+\frac{ B_0^4 \left(m-m_s\right){}^2 \left(19 m^2+mm_s+16m_s^2 \right)}{2187 \pi^2 M_\eta^4}
\nonumber\ena

$\bullet$ Diagrams 10
\bea
&&
-\frac{
  4 B_0^5  m^2 \left(m-m_s\right){}^3}{243 \pi^2 M_\eta^6}  \lnmpi
-\frac{ 4B_0^2  m m_s \left(m-m_s\right){}^3}{729 \pi^2 M_\eta^6}  \lnmk
\\
&&
-\frac{ 4B_0^5  \left(m-4 m_s\right)\left(m+8m_s\right)
\left(m-m_s\right){}^3}{6561 \pi^2  M_\eta^6}  \lnmeta\nonumber\\
&&
   -\frac{ 2B_0^5  (m +m_s) m_s \left(m-m_s\right){}^3}{243 \pi^2 M_\eta^6}  \lnmkkr
-\frac{4B_0^5 \left(m-m_s\right){}^3 \left(28 m^2+13 m m_s-32 m_s^2\right)}{6561 \pi^2 M_\eta^6}
\nonumber\ena
  
Let us denote $C(m,m_s)$ the sum of all these diagrams. One could naively think that $C$ is the NLO expression for $c_4$ once the propagators are reexpressed in terms of the physical $\eta$ mass, so that we just have to introduce a HO remainder $d_{c_4}^{\rm pole}$ of the form 
\begin{equation}
c_4^{\rm pole}=C+c_4^{\rm pole} d_{c_4}^{\rm pole}\,.
\end{equation}
However, like in the case of the topological susceptibility $\chi$~\cite{Bernard:2012fw}, one notices that $C$ does not vanish in the limit $m\to 0$ and/or $m_s\to 0$, contrary to the expectation that $c_4^{\rm pole}\to 0$ then, since the effective potential becomes independent of $\theta$ in these limits. This definition of $d_{c_4}^{\rm pole}$ must thus exhibit a divergent contribution in these chiral limits, so that $c_4^{\rm pole} d_{c_4}^{\rm pole}$ can compensate the non-vanishing value of $C$ to fulfill $c_4^{\rm pole}\to 0$.

Such a behaviour of $d_{c_4}^{\rm pole}$ is not very satisfying, as we would like HO remainders to become smaller and smaller in the limits of vanishing quark masses. This can be resolved by defining the remainder as
\begin{equation}\label{eq:c4NLOpole}
c_4^{\rm pole}(m,m_s)=C(m,m_s)-C(m,0)-C(0,m_s)+c_4^{\rm pole} d_{c_4}^{\rm pole}\,.
\end{equation}
This corresponds formally to a redefinition of the HO remainder, but has the important consequence that 
$d_{c_4}^{\rm pole}$ has no divergences and should remain small when approaching
in the two massless limits. This remainder can thus be approximated by a polynomial, and in the physical setting where $m_s$ is much larger than $m$, we expect $d_{c_4}^{\rm pole}=O(m_s^2)$.

The difference between the two expressions Eqs.~(\ref{eq:c4NLO2logs}) and (\ref{eq:c4NLOpole}) corresponds to a reshuffling between the sum of LO and NLO contributions and HO remainders. Since we hope to achieve a formulation with small HO remainders, the numerical differences between these expressions may become important to extract information on the pattern of $N=3$ chiral symmetry breaking
from precise lattice values of $c_4$~\footnote{Let us mention that in the simpler case of the topological susceptibility, the two formulations yield very comparable results in practice, as shown in Ref.~\cite{Bernard:2012fw}.}. For the moment, since no such precise determinations of $c_4$ are available, we will not pursue the comparison further, and we will restrict our discussion to 
Eq.~(\ref{eq:c4NLO}), bearing in mind that similar comments should certainly hold for the other formulations Eqs.~(\ref{eq:c4NLO2logs}) and (\ref{eq:c4NLOpole}).

\section{The cumulants of the topological charge and the pattern of chiral symmetry breaking}\label{sec:chicond}

\subsection{Expected sensitivity to the three-flavour chiral condensate}

We are now in a position to discuss the sensitivity of the cumulants of the topological charge to the 
pattern of chiral symmetry breaking in the limit of three massless quarks.
We see that the topological susceptibility Eq.~(\ref{eq:chit}) involves the three-flavour quark condensate
\begin{equation}
\Sigma(3)=-\lim_{m_u,m_d,m_s\to 0} \langle 0|\bar{u}u|0\rangle=F_0^2B_0\,,
\end{equation}
together with the Zweig-rule violating NLO LEC $L_6$ in the combination~\cite{Bernard:2012fw}
\begin{equation}
\chi \leftrightarrow F_0^2B_0 + 32 B_0^2L_6^r s\,,
\end{equation}
where $s=m_u+m_d+m_s$.
As discussed in Refs.~\cite{DescotesGenon:1999uh}, for physical quark masses, this combination is actually very close to the quark condensate defined in the two-flavour chiral limit $m_u,m_d\to 0$, but $m_s$ kept at its physical value
\begin{equation}
\Sigma(2)=-\lim_{m_u,m_d\to 0} \langle 0|\bar{u}u|0\rangle=
   F_0^2B_0  + 32m_s B_0^2 L_6^r(\mu) - \frac{m_sB_0^2}{16\pi^2}\log \frac{m_s B_0}{\mu^2}
   - \frac{m_s B_0^2}{72\pi^2}\log\frac{4m_s B_0}{3\mu^2} +\ldots
\end{equation}
where the ellipses denote HO contributions starting at  $O(m_s^2)$.
For simulations close to the physical case where $m$ is much smaller than $m_s$, the topological susceptibility essentially probes the two-flavour quark condensate (as the two quantities differ only by a contribution suppressed by a factor $m/m_s$), and consequently, $\chi$ is \emph{not} a very sensitive probe of the three-flavour quark condensate (defined in the corresponding chiral limit $m_u,m_d,m_s\to 0$), contrary to what is often asserted. Indeed, these two condensates can be significantly different
in the case of large contributions from sea-quark $s\bar{s}$ pairs (encoded in the Zweig-rule violating LEC $L_6$),
as first discussed in Ref.~\cite{DescotesGenon:1999uh}, and supported by several analyses 
($\pi K$ scattering~\cite{Buettiker:2003pp,DescotesGenon:2007ta}, scalar form factors~\cite{Moussallam:1999aq,Moussallam:2000zf,DescotesGenon:2000di,DescotesGenon:2000ct}, lattice data~\cite{Bernard:2010ex,Bernard:2012fw}). Therefore, a determination of ``the'' quark condensate from the topological susceptibility for lattice quark masses close to the physical ones can only be interpreted as a confirmation of our current knowledge on the pattern of $N=2$ chiral symmetry breaking probed experimentally through $\pi\pi$ (re)scattering  and through lattice simulations of the spectrum of pseudoscalar mesons. But this does not shed much light on the pattern of $N=3$ chiral symmetry breaking, and in particular on the size of the three-flavour condensate.

This feature can be seen most easily once the topological susceptibility is expressed in the framework of Resummed Chiral Perturbation Theory (Re$\chi$PT)~\cite{DescotesGenon:1999uh,DescotesGenon:2000di,DescotesGenon:2000ct,DescotesGenon:2002yv,DescotesGenon:2003cg,DescotesGenon:2007ta,Kolesar:2008jr,Bernard:2010ex,Bernard:2012fw}. Such a framework has been designed to cope with the fact that three flavor chiral series are not necessarily saturated by their leading-order term. Indeed if there are large contributions from sea-quark $s\bar{s}$ pairs and a suppression of the three-flavour condensate, a numerical competition between leading and next-to-leading orders will occur. In this case  an arbitrary function of pseudoscalar observables will not necessarily  have a well-convergent series  and one should not truncate chiral series by neglecting higher-order contributions. 

In that framework the topological susceptibility for physical quark masses 
is given by
\begin{equation}\label{eq:chinopolephysical}
\chi^{\rm no\ pole}_{\rm physical}=\frac{F_\pi^2M_\pi^2}{2}\frac{r}{2r+1}
 \Bigg[
   1-\frac{(7r+2)\epsilon(r)}{2r(2r+1)} 
+\frac{9}{2}\frac{r}{(r-1)^2(r+2)}\frac{3F_\eta^2M_\eta^2-4F_K^2M_K^2+F_\pi^2M_\pi^2}{F_\pi^2M_\pi^2}\Bigg]+ 
\ldots
\end{equation}
where $\epsilon( r)$ is a simple function of $r=m_s/m$ defined in Eq.~(\ref{funcr}), and the ellipsis denotes higher-order remainders starting at next-to-next-to-leading order (see Ref.~\cite{Bernard:2012fw} for a more extensive discussion of the topological susceptibility)~\footnote{We take this opportunity to point out that Eq.~(26) in Ref.~\cite{Bernard:2012fw} should be written with a factor 16 (rather than 4) in front of the term with $\hat{L}_6^r$ and $\hat{L}_8^r$, so that the last two terms in this equation cancel and yield Eq.~(\ref{eq:chinopolephysical}).}.
It is clear from this expression that $\chi$ is essentially sensitive to $r$, but exhibits no sensitivity to $\Sigma(3)$. 
This is related to the fact, shown in  Ref.~\cite{DescotesGenon:1999uh}, that the three-flavour chiral expansion of $F_\pi^2M_\pi^2$ yields a strong correlation between the two-flavour condensate and $r$.

If determining the three-flavour condensate from the topological susceptibility around the physical point seems very diffcult, the situation becomes more promising when turning to the fourth cumulant $c_4$, which involves $L_6$ in two different combinations
\begin{equation}
c_4 \leftrightarrow F_0^2B_0 + 32 B_0^2L_6^r s\ ,\qquad  -96 B_0^2 L_6^r \frac{\bar{m}^{[3]}}{\bar{m}^2}\,,
\end{equation}
For simulations close to the physical quark mass hierarchy, the first term corresponds 
 to the two-flavour condensate (like for the topological susceptibility), but
the second term opens the possibility to disentangle $L_6^r$  and $\Sigma(3)$ by considering the dependence of $c_4$ on the light-quark mass $m$.  Therefore,
contrary to the topological susceptibility $\chi$, the fourth cumulant $c_4$ may provide a direct probe of the pattern of three-flavour chiral symmetry breaking even for simulations close to the physical case.

Before turning to the numerical analysis of lattice settings close to the physical quark mass hierarchy, let us emphasise that simulations with degenerate quarks could be much more interesting to extract the three-flavour condensate from topological quantities. Indeed, in the limit where the $N$ quarks in Eqs~(\ref{eq:chit}) and (\ref{eq:c4}) are degenerate with mass $\hat{m}$ (leading to pseudoscalar mesons with a common physical mass $M$), the following combination
\begin{equation}\label{eq:deg}
\chi^{\rm no\ pole}+\frac{N^2}{4}c_4^{\rm no\ pole}=\frac{3\hat{m}^2F_N^2B_N}{4N}+\frac{N^2-1}{N^2}\frac{3\hat{m}^2B_N^2}{32\pi^2}
 -\frac{N^2-1}{N^2}\frac{3\hat{m}^2B_N^2}{32\pi^2}\log\frac{2\hat{m}B_N}{M^2} +O(\hat{m}^3)\,,
\end{equation}
does not involve NLO LECs, and is particularly suited to determine the quark condensate $\Sigma(N)=F^2_N B_N$ in the chiral limit of $N$ massless flavours from simulations with $N$ degenerate light quarks.

\subsection{Numerical analysis}

Several lattice computations of the (normalised) fourth cumulant have been performed in pure gauge theory for different number of colours~\cite{DelDebbio:2002xa,D'Elia:2003gr,Giusti:2007tu}. Even though their values for 3 colours are rather close to those that we will obtain in the following, they cannot be used to investigate the role of strange sea-quark pairs in chiral symmetry breaking (see Ref.~\cite{Vicari:2008jw} for more details and references on these simulations). There exist also computations with 2 and (2+1) dynamical fermions from the TWQCD collaboration~\cite{Chiu:2011dz,Chiu:2008jq}, whose central values are in the same ball park as our results, but with so large uncertainties that any practical comparison with our figures is meaningless at the current stage. These computations are therefore at an earlier stage than the well-studied case of the topological susceptibility~\cite{Durr:2001ty,
Giusti:2004qd,DelDebbio:2004ns,Durr:2006ky,Bernard:2007ez,Aoki:2007pw,Chiu:2008kt,Vicari:2008jw,Bazavov:2010xr,Chowdhury:2011yj,Cichy:2011an,Chiu:2011za}.

Our analysis will thus be limited to a proof of principle that the fourth cumulant $c_4$ is indeed an interesting probe of the pattern of three-flavour chiral symmetry breaking. 
We consider Eq.~(\ref{eq:c4NLO2logs}) in the case of lattice simulations with (2+1) dynamical quarks of masses $\tilde{m}$ and $\tilde{m}_s$ measured as
\begin{equation}\label{eq:pq}
p=\frac{\tilde{m}_s}{m_s} \,,\qquad q=\frac{\tilde{m}}{\tilde{m}_s}\,.
\end{equation}
We introduce the parameters involved in the LO chiral Lagrangian
\begin{equation}\label{eq:LOLECs}
X(3)=\frac{2m\Sigma(3)}{F_\pi^2M_\pi^2}\,,\qquad Z(3)=\frac{F^2(3)}{F_\pi^2}\,,\qquad r=\frac{m_s}{m}\,,
\end{equation}
One can also introduce the ratio of order parameters
\begin{equation}
Y(3)=\frac{X(3)}{Z(3)}=\frac{2m B_0}{M_\pi^2}\,.
\end{equation}
$X(3)$, $Y(3)$ and $Z(3)$ indicate the saturation of the three-flavour chiral expansions of $F_\pi^2M_\pi^2$, $M_\pi^2$ and $F_\pi^2$ by their leading-order contribution respectively. As indicated before, there are several indications that $X(3)$ and $Z(3)$ are not particularly close to 1, and that the three-flavour chiral expansions of pseudoscalar masses and decay constants experience a numerical competition between leading and next-to-leading orders.

Denoting $\tilde{X}$ an observable measured on the lattice, and $X$ the same observable at the physical point, we obtain for the fourth cumulant on the lattice
\bea
&&
\tilde{c}_4^{\rm no\ pole}=-\frac{{\fpid} \mpid p q \left(q^3+2\right) r {X(3)}}{2 (q+2)^4}+\frac{{\mpiq} p^2 q^2 
( (13 q^2+22q+37) r^2 {Y(3)}^2}{96 \pi ^2
   (q+2)^4} 
\\
&&
-\frac{16 L_6^r(\mu) \mpiq p^2 q \left(q^4+2 q^3+6 q^2+8 q+1\right) r^2 {Y(3)}^2}{(q+2)^4}
-\frac{96 (3L_7+L_8^r(\mu)) \mpiq p^2 q^2
   \left(q^3+2\right) r^2 {Y(3)}^2}{(q+2)^5}
\nonumber\\
&&
+\frac{9 {\mpiq} p^2 q^2 r^2 {Y(3)}^2}{32 \pi ^2 (q+2)^4} \lnmpil
+\frac{3 {\mpiq} p^2 q^2 \left(4 q^3-3
   q+2\right) r^2 {Y(3)}^2}{32 \pi ^2 (q+2)^5} \lnmpikrl
\nonumber\\
&&
-\frac{ \mpiq 3 p^2 q^2 (q+1)^2 r^2 {Y(3)}^2}{32 \pi ^2 
(q+2)^4} \lnmkl 
+\frac{
 \mpiq p^2 q (q+1)^2 \left(q^3-3q^2+3q+2\right) r^2 {Y(3)}^2}
{32 \pi ^2 (q+2)^5} \lnmkkrl
\nonumber\\
&&
+\frac{ \mpiq p^2 q^2 (2 q+1)^2 r^2 {Y(3)}^2}{96 \pi ^2
   (q+2)^4} \lnmetal
+\frac{ \mpiq p^2 q \left(2q^4-8q^3+13q+2\right) r^2 {Y(3)}^2}{288 \pi ^2
   (q+2)^4} \lnmetakrl + \tilde{c}_4^{\rm no\ pole} {\tilde d}_{c_4}^{\rm no\ pole}\nonumber
    \label{eq:totnopole}\,.
\ena
In Resummed $\chi$PT~\cite{DescotesGenon:2003cg,DescotesGenon:2007ta,Bernard:2010ex}, $L_{6,7,8}^r(\mu)$ are replaced by their expressions derived from the chiral expansion of observables expected to have good convergence properties ($F_P^2$ and $F_P^2 M_P^2$ with $P=\pi,K,\eta$). The corresponding formulae were given in refs~\cite{Bernard:2010ex,Bernard:2012fw} and are recalled in App~\ref{app:L6L7L8}. $\tilde{d}_{c_4}$ is the HO remainder which is set to zero in the present analysis (it is expected to be of order $O(\tilde{m}_s^2)$ and thus with the scaling $\tilde{d}_{c_4}\simeq p^2 d_{c_4}$ for simulations where $q=\tilde{m}/\tilde{m_s}$ is small). 

\begin{figure}[t!]
\begin{center}
\includegraphics[width=12cm,angle=0]{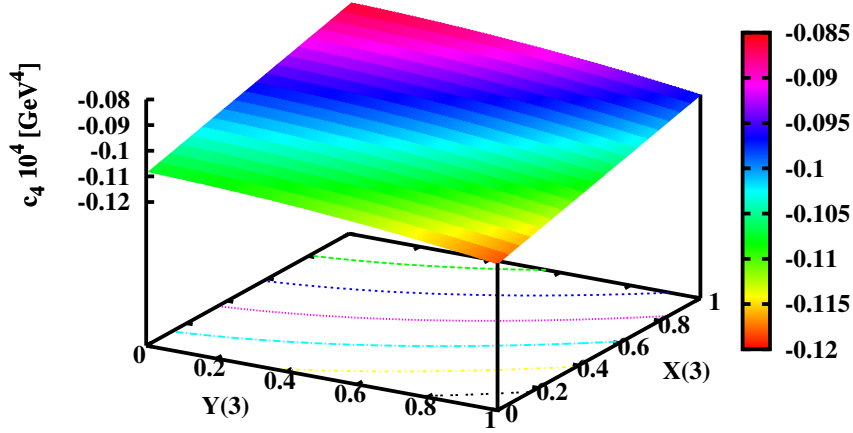}
\caption{\small\it $c_4$ (in units of $10^4$ GeV$^4$) as a function of $X(3)$ and $Y(3)$ for $p=1,r=1/q=25$, and all HO remainders set to zero. \label{figure:c4}}
\end{center}
\end{figure}

The lattice counterpart of Eq.~(\ref{eq:c4NLO})  can be obtained by replacing all masses in the logarithms by their LO values denoted as  $\klmpip$, yielding
\begin{eqnarray}\label{eq:totnopolebis}
&&
\tilde{c}_4^{\rm no\ pole}=-\frac{{\fpid} \mpid p q \left(q^3+2\right) r {X(3)}}{2 (q+2)^4}+\frac{{\mpiq} p^2 q^2 
( (13 q^2+22q+37) r^2 {Y(3)}^2}{96 \pi ^2
   (q+2)^4} 
\\
&&
-\frac{16 L_6^r(\mu) \mpiq p^2 q \left(q^4+2 q^3+6 q^2+8 q+1\right) r^2 {Y(3)}^2}{(q+2)^4}
-\frac{96 (3L_7+L_8^r(\mu)) \mpiq p^2 q^2
   \left(q^3+2\right) r^2 {Y(3)}^2}{(q+2)^5}
\nonumber\\
&&
+\frac{3 {\mpiq} p^2 q^2 \left(2+q^2\right) r^2 {Y(3)}^2}{8 \pi ^2 (q+2)^5} \lnmpikrl
+\frac{
 \mpiq p^2 q (q+1)^2 \left(q^3+9q+2\right) r^2 {Y(3)}^2}
{32 \pi ^2 (q+2)^5} \lnmkkrl
\nonumber\\
&&
+\frac{ \mpiq p^2 q \left(q^4+2q^3+6q^2+8q+1\right) r^2 {Y(3)}^2}{144 \pi ^2
   (q+2)^4} \lnmetakrl + \tilde{c}_4^{\rm no\ pole} {\tilde d}_{c_4}^{\rm no\ pole}
\nonumber\,.
\end{eqnarray}

\begin{figure}[t!]
\begin{center}
\includegraphics[width=8.3cm,angle=0]{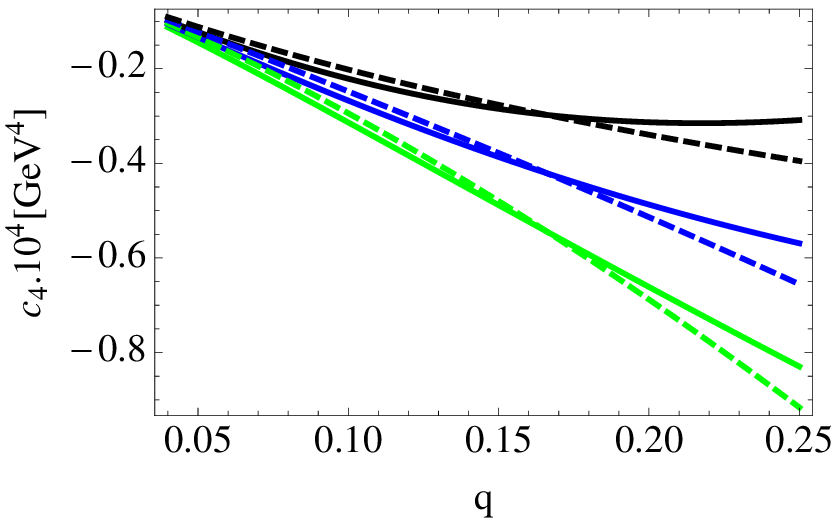}
\includegraphics[width=8.3cm,angle=0]{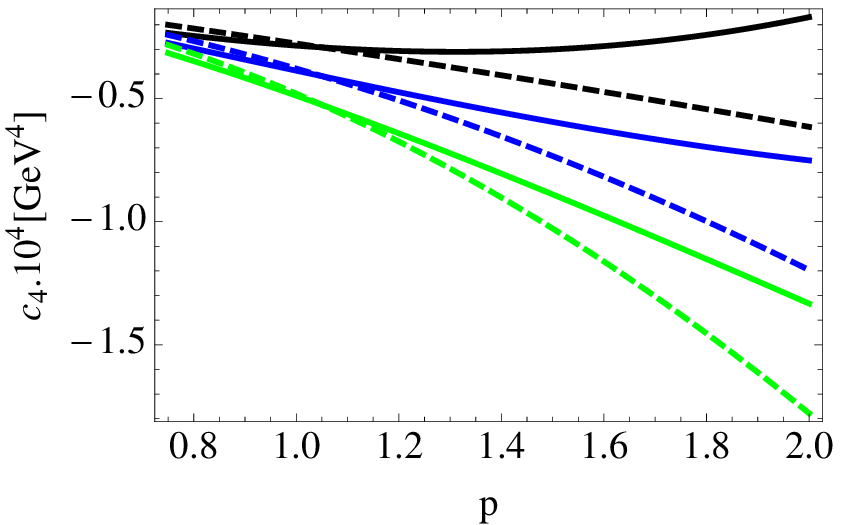}
\caption{\small\it $c_4$ as a function of $q=\tilde{m}/\tilde{m}_s$ for $p=1$ (left panel) and as a function of $p=\tilde{m}_s/m_s$ for $q=0.15$ (right panel). Light (green), dark (blue) and black curves correspond to $X(3)=0.2,0.5,0.8$ respectively,
whereas solid and dashed curves correspond to $Y(3)=0.9$ and 0.2 respectively. We take $r=25$ and all HO remainders are set to zero.
\label{figure:c4wrtq}}
\end{center}
\end{figure}

\begin{figure}[t!]
\includegraphics[width=8.25cm,angle=0]{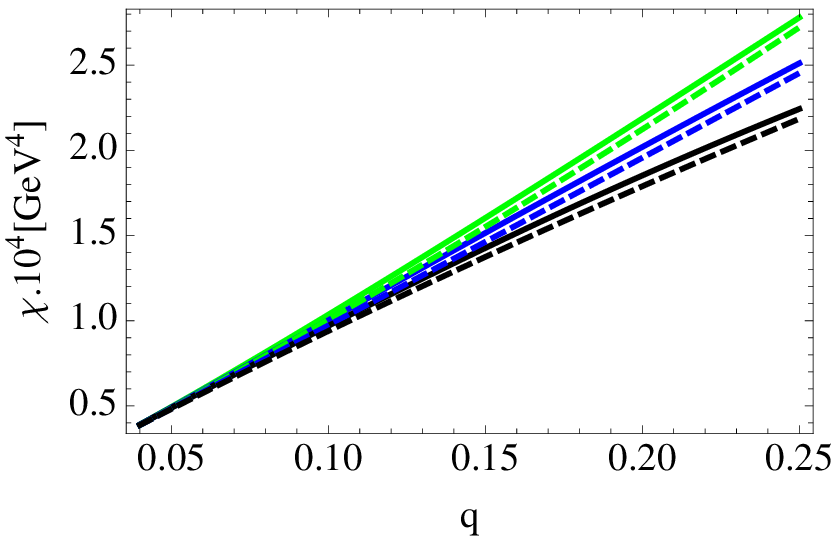}
\includegraphics[width=8.5cm,angle=0]{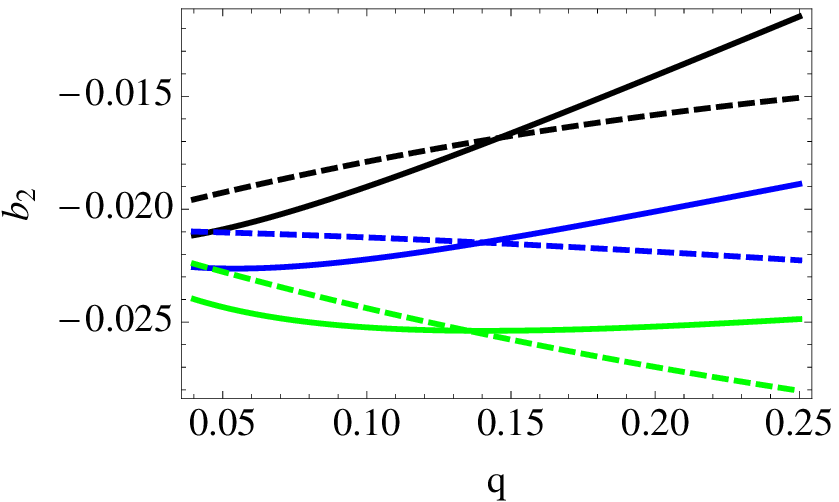}
\caption{\small\it $\chi$ (left panel) and $b_2=c_4/(12\chi)$ (right panel) as a function of $q=\tilde{m}/\tilde{m_s}$
for $p=1,r=25$. Light (green), dark (blue) and black curves correspond to $X(3)=0.2,0.5,0.8$ respectively,
whereas solid and dashed curves correspond to $Y(3)=0.9$ and 0.2 respectively. All HO remainders are set to zero.
\label{figure:chib2}}
\end{figure}

\begin{figure}[t!]
\begin{center}
\includegraphics[width=8.2cm,angle=0]{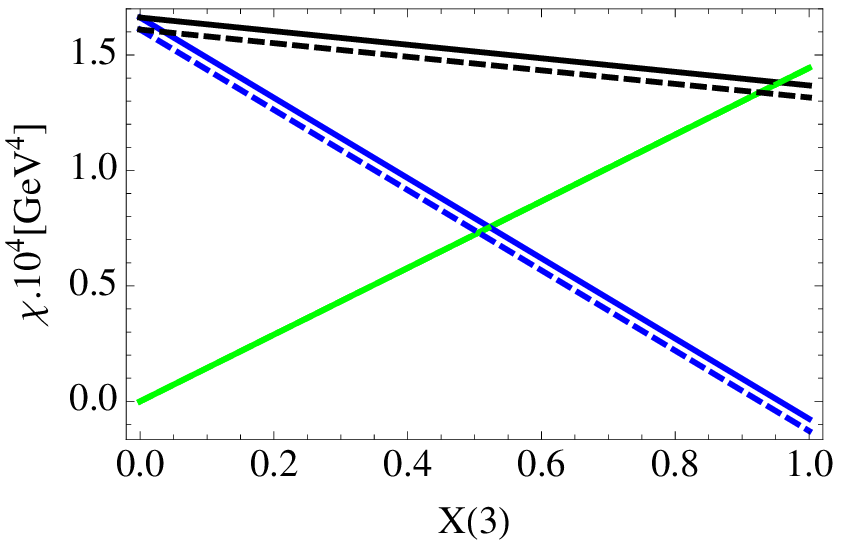}
\includegraphics[width=8.45cm,angle=0]{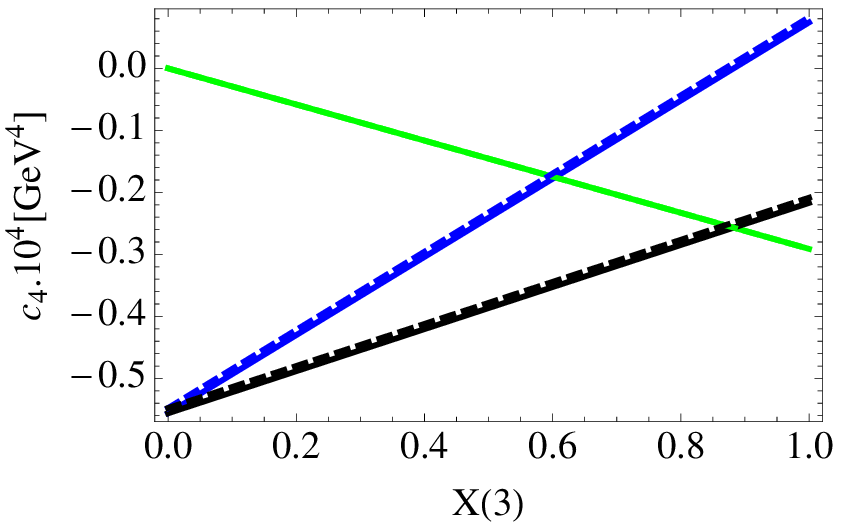}
\end{center}
\caption{\small\it Order-by-order contributions to $\chi$ (left panel) and $c_4$ (right panel) as a function of $X(3)$ in the case $p=1,q=0.15,r=25$. Light (green), dark (blue)  and black curves correspond to LO contributions, NLO ones  and their sum respectively. Solid and dashed curves correspond to $Y(3)=0.9$ and 0.2 respectively (they are identical for the LO contribution which depends on $X(3)$ and quark masses only). All HO remainders are set to zero.}
\label{figure:c4conv}
\end{figure}

We will use Eq.~(\ref{eq:totnopolebis}) to determine the potentiality of the fourth cumulant $c_4$ to extract the three-flavour quark condensate from simulations performed with a mass hierarchy similar to the physical situation (for this first numerical investigation, 
we use $M_\pi = 139.6$ MeV, $M_K = 495.7$ MeV, $M_\eta = 547.8$ MeV,
$F_\pi=92.2$ MeV, $F_K/F_\pi = 1.18$).
All higher-order (HO) remainders  are set to zero.

We first consider the physical case where $p=1,q=1/r$, and in Fig.~\ref{figure:c4},  we show
the dependence of $c_4$ on $X(3)$ and $Y(3)$ assuming $r=25$. We see a noticeable sensitivity to $X(3)$ and a weak sensitivity to $Y(3)$, as expected from the discussion in Sec.~\ref{sec:chicond}. 
Let us move from the physical case to consider simulations with a near physical strange quark mass and light quark masses larger than the physical one (i.e., $p$ close to 1 and $q\geq 1/r$). The variation of $c_4$ with respect to $p$ and $q$, $X(3)$ and $Y(3)$ is shown in Fig.~\ref{figure:c4wrtq}. The slope of the curve is very dependent on $X(3)$, and more weakly on $Y(3)$, and the dependence of $c_4$  on the light-quark mass $\tilde{m}$ provides an efficient probe of the three-flavour quark condensate $X(3)$. 

As indicated in the qualitative discussion of the previous section, the topological susceptibility $\chi$
has a much weaker sensitivity to $X(3)$, which arises only for values of $q$ closer to 1, i.e., light quark masses away from the physical situation. Therefore, a meaningful extraction of the three-flavour quark condensate is possible only for 
unphysical quark mass hierarchies.
This can be seen from the left-hand side of Fig.~\ref{figure:chib2} (we plot here Eq.~(\ref{eq:chinopole}), obtained without singling out the $\eta$-pole and denoted $\chi^{\rm no\ pole}$ in Ref.~\cite{Bernard:2012fw}).  The relative sensitivities of
$\chi$ and $c_4$ on the three-flavour quark condensate have however to be balanced with the accuracy that can be expected from lattice simulations. As discussed for instance in ref.~\cite{Durr:2006ky}, the fourth cumulant is much more challenging to determine through lattice simulations than the topological susceptibility, due to the higher order of the correlator making the extraction of the signal quite difficult. In the coming years, it appears thus likely that the measurement of both quantities will provide only limited information on the three-flavour quark condensate. 

We also display the normalised fourth cumulant $b_2=c_4/(12\chi)$ on the right hand-side  of Fig.~\ref{figure:chib2}, which exhibits the same appealing sensitivity to the three-flavour quark condensate  as $c_4$. We
notice that its dependence on the simulated light quark mass is modified compared to the LO result~\cite{Mao:2009sy}
\begin{equation}
\tilde{b}_{2;LO}=-\frac{2+q^3}{12(2+q)^3}=-\frac{1}{48}+\frac{q}{32}+O(q^2)\,,
\end{equation}
both for the value at the origin and the slope when the quark condensate does not saturate the thee-flavour expansions of quark masses. Different patterns of three-flavour chiral symmetry breaking, and in particular, different values of the three-flavour quark condensate, can modify the dependence of this cumulant on the simulated light quark mass in a striking way.

In Fig.~\ref{figure:c4conv}, we indicate the LO and NLO contributions to $\chi$ and $c_4$ for a simulation with $p=1,q=0.15$ and $r=1/q$. A significant cancellation of the two contributions in the case of $c_4$, emphasising the need to  perform the analysis of this quantity in $\chi$PT up to one loop, whatever the pattern of chiral symmetry breaking. Moreover, this partial cancellation does not prevent the sum of LO and NLO contributions for $c_4$ to exhibit a strong dependence on $X(3)$, contrary to the topological susceptibility $\chi$ which again shows a very limited sensitivity to $X(3)$ (as expected since $q=\tilde{m}/\tilde{m}_s$ is small).

\begin{table}[t!]
$$
\begin{array}{|c|c|}
\hline
r       &  23.7\pm 0.5 \\
X(3)     & 0.38\pm 0.05\\
Y(3)     & 0.71\pm 0.10\\
Z(3)    & 0.54\pm 0.06 \\
F_K/F_\pi & 1.17\pm 0.01 \\

\phantom{xx}&\\[-2.9ex]
m_s(2\ {\rm GeV}) [{\rm MeV}]  & 92.3\pm 3.7 \\
m(2\ {\rm GeV})   [{\rm MeV}]  & 3.9\pm 0.2  \\
\Sigma_0^{1/3}(2\ {\rm MeV}) [{\rm GeV}]  & 201\pm 9 \\
B_0(2\ {\rm GeV}) [{\rm GeV}] & 1.77\pm 0.27\\
F_0 [{\rm MeV}]               & 67.6\pm 3.4\\
\hline
\phantom{xx}&\\[-2.9ex]
c_1 & 5.53 \pm 0.75\\
c_2 & 1.53 \pm 0.42\\
c_{F_\pi} [{\rm GeV}^2]& 0.02 \pm 0.10\\
c_{F_K} [{\rm GeV}^2] & 0.20 \pm 0.16\\
\hline
\phantom{xx}&\\[-2.9ex]
 L_4(M_\rho)\cdot 10^3    & 1.06\pm 0.29 \\
 L_5(M_\rho)\cdot 10^3    & 1.48\pm 0.22\\
 L_6(M_\rho)\cdot 10^3    & 1.19\pm 0.41 \\
 L_7(M_\rho)\cdot 10^3    & -1.06\pm 0.37\\
 L_8(M_\rho)\cdot 10^3    & 1.64\pm 0.60\\
\hline
\phantom{xx}&\\[-2.9ex]
\Sigma/\Sigma_0 & 2.36 \pm 0.29\\
B/B_0 & 1.42 \pm 0.17\\
F/F_0 & 1.29 \pm 0.07
 \\[0.15ex]
 \hline
\end{array}
$$
\caption{\small\it Updated results using the same approach as in ref.~\cite{Bernard:2012fw}, through a fit to the RBC/UKQCD data on pseudoscalar masses, decay constants and topological susceptibility~\cite{Aoki:2010dy,:2012he} within the framework of Resummed Chiral Perturbation Theory.
\label{table:num}
}
\end{table}

\begin{table}[t!]
\begin{center}
\begin{tabular}{|c|cc|ccc|}
\hline
Case & $p$ & $q$ & $\chi\cdot 10^4$ [GeV$^4$] &  $c_4\cdot 10^4$ [GeV$^4$] & $b_2$\\
\hline
$L_s=24$ & 1.22 & 0.192 &  $3.53 \pm 0.13$ & $-0.68\pm 0.12$ & $-0.016\pm 0.003$\\
 & 1.22 & 0.311 & $5.64\pm 0.31$ & $-1.07\pm 0.25$ &  $-0.016\pm  0.003$\\
 \hline
$L_s=32$ & 1.18 & 0.154 & $2.62\pm 0.08$ & $-0.52\pm 0.09$ & $-0.017 \pm 0.003$\\
  & 1.18 & 0.220 &  $3.73\pm 0.14$ & $-0.73\pm 0.15$ & $-0.016\pm 0.003$\\
 & 1.18 & 0.286 & $4.82\pm 0.23$ & $-0.94\pm 0.21$ & $-0.016\pm 0.003$\\
 \hline
 $L_s=32$ DSDR & 1.06 & 0.061 & $ 0.84\pm 0.03$ & $-0.18\pm 0.03$ & $-0.018 \pm 0.003$\\
  & 1.06 & 0.131 &  $1.79\pm 0.05$ & $-0.38\pm 0.07$ & $-0.017\pm 0.003$\\
 \hline
Physical & 1 & $1/r$ & $0.51\pm 0.01$ & $-0.11\pm 0.03$ &$-0.018\pm 0.003$\\
\hline
\end{tabular}
\caption{{\small \it Predictions for the topological susceptibility and the fourth cumulant (without identifying the $\eta$ pole) for different lattice data sets 
produced by the RBC and UKQCD collaborations~\cite{Aoki:2010dy,:2012he} and in the physical case, following the analysis performed in Ref.~\cite{Bernard:2012fw} leading to the results summarised in Table 1.}} \label{tab:pred}
\end{center}
\end{table}

Finally, we illustrate typical values for the fourth cumulant $c_4$ and $b_2$ for current lattice simulations. We consider the update of fit  B5 from Ref.~\cite{Bernard:2012fw}, corresponding to the analysis of the spectrum of light pseudoscalar mesons and the topological susceptibility from the RBC and UKQCD collaborations~\cite{Aoki:2010dy} within the framework of Resummed $\chi$PT~\cite{DescotesGenon:2003cg,DescotesGenon:2007ta,Bernard:2010ex}. We update this analysis by including the new Iwasaki DSDR data set presented in ref.~\cite{:2012he}~\footnote{This update has confirmed the analysis of ref.~\cite{Bernard:2012fw} concerning the difficulties of a precise determination of lattice spacings, which plays an essential role in setting the absolute scale for the measured observables on the lattice (decay constants, masses), but not on the discretisation effects, which were found to be negligible in ref.~\cite{Aoki:2010dy} (we take this opportunity to correct the false statement of ref.~\cite{Bernard:2012fw}, attributing large discretisation effects in ref.~\cite{Aoki:2010dy}).}. This analysis leads to
the values gathered in Table~\ref{table:num}, with a very good minimum $\chi^2$ at 22.2 for 19 parameters and 39 points, corresponding to 1.1 $\sigma$ if all uncertainties are interpreted as Gaussian. The values  of the quark condensate 
and the pseudoscalar decay constant in the three-flavour chiral limit are similar to those in ref.~\cite{Bernard:2012fw}, as well as the constants describing discretisation artefacts, but the lattice spacings are found closer to the estimates from ref.~\cite{Aoki:2010dy} (our values are actually 4\% lower for all three data sets). This induces noticeable changes for the quantities with dimensions compared to ref.~\cite{Bernard:2012fw}, but the overall pattern of chiral symmetry breaking remains is very similar.

Using the outcome of this fit (including the correlations among the LO order parameters and the HO remainders), we obtain the predictions for $\chi$, $c_4$ and $b_2$ collected in Table~\ref{tab:pred} for the different sets considered in Ref.~\cite{Bernard:2012fw} as well as in the physical case. The negative value of $b_2$ indicates that the distribution of the topological charge is flatter than a pure Gaussian.
The values for $p=\tilde{m}_s/m_s$ 
are different from those quoted in Ref.~\cite{Aoki:2010dy}, since we reassessed the determination of the lattice spacing and quark mass based on the mass of the $\Omega$ baryon in ref.~\cite{Bernard:2012fw}. Finite-volume effects are not included, whereas the effect of HO remainders is taken into account through the fit  and an additional uncertainty coming from $\tilde{d}_{c_4}\simeq 0.15$ has been added to the results for $\tilde{c}_4$ and $\tilde{b}_2$. As in the case of the topological susceptibility illustrated in Ref.~\cite{Bernard:2012fw}, the impact of  $\tilde{d}_{c_4}$ can be easily modeled and determined from a fit once a sufficiently large set of values at different simulated quark masses is available. Obviously, it would be very interesting to compare these predictions for $c_4$ and $b_2$ with data from lattice simulations.

\section{The two-flavour case}\label{sec:twoflav}

As indicated after Eq.~(\ref{eq:c4}), it is possible to compute the topological susceptibility or the fourth cumulant using the $N=2$ $\chi$PT Lagragian described in Ref.~\cite{Gasser:1983yg} and built around the two-flavour chiral limit $m_u=m_d=0$ (but $m_s$ kept physical).
 The only degrees of freedom of the theory are soft pions and the resulting expansions are series in powers of $m_u$ and $m_d$ only. The LECs $\ell_i$ and $h_i$ involved are by construction
different from those in $N=3$ $\chi$PT (denoted $L_i$ and $H_i$), as they are defined in two distinct chiral limits. 
For instance, at leading order, two-flavour $\chi$PT will involve the pseudoscalar decay constant $F$ and the quark condensate $-F^2B$ defined in the limit $m_u=m_d=0$, but $m_s$ physical, and one has
\begin{equation}
\lim_{m_s\to 0} F^2 = F_0^2\,, \qquad \lim_{m_s\to 0} -F^2B = -F_0^2B_0\,,
\end{equation}
showing that the two sets of quantities are in principle different: one expects
that two-flavour chiral order parameters are larger than the three-flavour ones~\cite{DescotesGenon:1999uh},
and our fits to lattice data indeed confirm this trend with a significant suppression of chiral order parameters when the strange quark mass is sent to zero~\cite{Bernard:2010ex,Bernard:2012fw}. 

A further complication stems from the fact that the structure of NLO chiral Lagrangian used in Refs.~\cite{Gasser:1984gg} and \cite{Gasser:1983yg} is different, so that the correspondence between the LECs in the two theories is not immediate beyond leading order.
It is straightforward to repeat the same arguments as in Sec.~\ref{sec:genfunc} with
the generating functional of  $N=2$ $\chi$PT given in Ref.~\cite{Gasser:1983yg}, in order to obtain the two-flavour expressions in the isospin limit
\begin{eqnarray}\label{eq:chitwo}
\chi^{\rm no\ pole}&=&\frac{mBF^2}{2} + 2m^2B^2 [\ell_3^r(\mu) + h_1^r(\mu) - \ell_7 - h_3]
           - \frac{3B^2m^2}{32\pi^2} \log\frac{2mB}{\mu^2}+O(m^3)\\
c_4^{\rm no\ pole}&=&-\frac{mBF^2}{8} + \frac{9m^2B^2}{128 \pi^2}
           - 2m^2B^2 [\ell_3^r(\mu) + h_1^r(\mu) - \ell_7 - h_3]\\
&&       \qquad    + \frac{3m^2B^2}{128\pi^2} \left[\log\frac{2mB}{\mu^2}+3 \log\frac{M_\pi^2}{\mu^2}\right]+O(m^3)\label{eq:c4two}\nonumber\,,
\end{eqnarray}
where we have performed the same separation between unitarity and tadpole contributions for $c_4$ as in the $N=3$ case.
These formulae should be used to extract the two-flavour chiral condensate from simulations performed at $\tilde{m}_s=m_s$ and various values of the light quark mass $\tilde{m}$.  One notices in particular that the sum of Eqs.~(\ref{eq:chitwo})-(\ref{eq:c4two}), i.e. $c_4+\chi$, yields a result in agreement with Eq.~(\ref{eq:deg}): the cancellation of all NLO LECs shows that this sum
would be a particularly clean probe of the two-flavour condensate $-BF^2$ in the isospin limit.

One can go one step further and actually match $N=2$ and $N=3$ chiral theories
in order to exhibit the $m_s$-dependence of the $N=2$ LECs~\cite{Gasser:1983yg,Gasser:2007sg,Gasser:2009hr,Gasser:2010zz}, yielding the NLO  matching formulae for the quark condensates and pseudoscalar decay constants
\begin{eqnarray}\label{eq:decmatch}
F^2&=& F^2_0 +16m_s B_0 L_4^r(\mu)- \frac{m_sB_0}{16\pi^2}\log \frac{m_s B_0}{\mu^2}\\
F^2B&=& F^2_0 B_0 + 32m_s B_0^2 L_6^r(\mu) - \frac{m_sB_0^2}{16\pi^2}\log \frac{m_s B_0}{\mu^2}
   - \frac{m_s B_0^2}{72\pi^2}\log\frac{4m_s B_0}{3\mu^2} \,,
\label{eq:condmatch}
\end{eqnarray}
as well as for the relevant NLO LECs
\begin{eqnarray}\label{eq:matchLECs}
2m^2B^2[\ell_3^r(\mu) + h_1^r(\mu) - \ell_7 - h_3] &=&
 24 m^2 B_0^2 [L_6^r+3L_7^r+L_8^r]
  -\frac{m^2B_0^2}{32\pi^2} \log\frac{m_s B_0}{\mu^2}\\
&&\qquad  -\frac{m^2B_0^2}{96\pi^2} \log\frac{4m_s B_0}{3\mu^2}
   -\frac{5m^2B_0^2}{144\pi^2} -\frac{F_0^2B_0m^2}{4 m_s} \nonumber\,,
\end{eqnarray}
where $L_{6,7,8}^r=L_{6,7,8;N=3}^r$ denote the LECs from the $N=3$ chiral Lagrangian.
It is easy to check that Eqs.~(\ref{eq:chitwo})-(\ref{eq:c4two}) correspond indeed to the second-order expansion in $m$ of the three-flavour results Eqs.~(\ref{eq:chinopole}) and (\ref{eq:c4NLO2logs}), providing a further cross-check of these equations.

At first sight, one might be surprised that two high-energy counterterms $h_1$ and $h_3$ arise in these expansions. Indeed, these high-energy counterterms encode physics of higher energy scales that are not included dynamically in the theory and their value is dependent on its ultraviolet regularisation. This seems to contradict the fact that $\chi$ and $c_4$ are topological quantities related to chiral symmetry breaking arising in QCD at low energies. This paradox can be solved thanks to the matching with three-flavour $\chi$PT. The two-flavour chiral expansions of $\chi$ and $c_4$ involve only the difference of high energy counterterms $h_1-h_3$, which can be matched with three-flavour $\chi$PT as in
Eq.~(\ref{eq:matchLECs}). The NLO expansion of $h_1-h_3$ in powers of $m_s$ involves only $N=3$ low-energy constants, but none of the high-energy counterterms $H_i$. It means that $h_1-h_3$ is a combination of $N=2$ high-energy counterterms which is characterised by the dynamics of $K$ and $\eta$ mesons but is independent of the ultraviolet regularisation of the theory in relation with more massive (non-Goldstone) degrees of freedom. Indeed we have seen that our three-flavour results for $\chi$ and $c_4$ rely crucially on the propagation of the $\eta$-meson, which is not a dynamical degree of freedom of two-flavour $\chi$PT. Therefore, all the diagrams in $N=3$ $\chi$PT involving the propagation of $K$ and $\eta$ mesons (such as Fig.~\ref{figure:LOdiag}) must be absorbed into high-energy counterterms once the computation is performed in $N=2$ $\chi$PT. The presence of $h_1-h_3$ in Eq.~(\ref{eq:decmatch}) is thus normal and can be easily explained by the peculiar role of the $\eta$ meson in  the NLO expansions of $\chi$ and $c_4$.

Finally, let us stress the ambiguity of the alternative  expression  used in the literature~\cite{Aoki:2010dy,Mao:2009sy} and derived from Eqs.~(\ref{eq:chit}) and (\ref{eq:c4}) by setting $N=2$. The resulting expressions have the same structure as Eqs.~(\ref{eq:chitwo}) and (\ref{eq:c4two}), but they involve LECs labeled $L_6,L_7,L_8$ that are very easy to misunderstand. Their names allude the three-flavour case, but they are actually defined in the two-flavour chiral limit. In our notation, we would denote them as $L_{6,7,8;N=2}$ which have definitions and values that differ from those in three-flavour $\chi$PT (which we would denote $L_{6,7,8;N=3}$). The connection between these LECs $L_{6,7,8;N=2}$  and the usual $\ell_i$ and $h_i$ can be obtained from refs.~\cite{Gasser:1983yg,Bijnens:1999hw,Bijnens:2009qm}\begin{eqnarray}
\ell_3^r&=&-8L_{4;N=2}^r-4L_{5;N=2}^r+16L_{6;N=2}^r+8L_{8;N=2}\,,\\
\ell_7&=&-16L_{7;N=2}-8L_{8;N=2}\,,\\
h_1^r&=&8L_{4;N=2}^r+4L_{5;N=2}^r-4L_{8;N=2}+2H_{2;N=2}\,,\\
h_3^r&=&4L_{8;N=2}^r+2H_{2;N=2}\,.
\end{eqnarray}
In order to avoid any confusion and make a direct link with phenomenological analyses performed in $N=2$ $\chi$PT~\cite{Gasser:1983yg}, one should always use the expressions Eqs.~(\ref{eq:chitwo}) and (\ref{eq:c4two}) to deal with the two-flavour chiral expansions of the topological susceptibility and the fourth cumulant.

\section{Conclusion}

Due to their connection with the $U_A(1)$ anomaly, topological observables describing the distribution of the topological charge (or gluonic winding number) are able to probe the structure of low-energy QCD, and in particular the pattern of chiral symmetry breaking. Moreover, they can be accessed through lattice QCD simulations and are thus complementary to the analysis of the quark mass dependence of observables related to light pseudoscalar mesons. The  two most prominent topological observables are the topological susceptibility $\chi$ and the fourth cumulant $c_4$, corresponding to the first two non-trivial moments of the distribution of the topological charge. The present paper is a follow-up of the investigation started in Ref.~\cite{Bernard:2012fw}, in order to determine if these observables can be used to constrain the pattern of three-flavour chiral symmetry breaking (defined in the $m_u=m_d=m_s$ chiral limit). We have performed the computation of the two quantities up to next-to-leading order in three-flavour Chiral Perturbation Theory using two methods: the elegant approach of the effective potential and the more pedestrian computation of Feynman diagrams.
In the first case, we derived expressions valid for an arbitrary number of light flavours, whereas in the second case, we were able to pin down the role played by the propagation of flavour-singlet mesons. This allowed us to emphasise the importance of reexpressing the denominators arising at next-to-leading order expression and involving the leading-order contribution of the $\eta$-mass in terms of its physical value.

We then investigated the sensitivity of the next-to-leading order expressions of the topological susceptibility and the fourth cumulant to the pattern of three-flavour chiral symmetry breaking in the context of lattice QCD simulations. We confirmed the result of Ref.~\cite{Bernard:2012fw} that for light-quark mass hierarchy close to the physical case, the topological susceptibility has mainly a sensitivity to the quark condensate in the two-flavour chiral limit ($m_u=m_d=0$ but $m_s$ physical), but not to the three-flavour one. Since there are phenomenological indications that the two condensates are significantly different, one has to devise alternative ways of obtaining the three-flavour quark condensate from topological observables. 

A possibility would consist in considering the topological susceptibility far from the physical situation, with values of $m_u,m_d,m_s$ much closer to each other. A novel alternative is provided by the fourth cumulant, which has indeed a much better sensitivity to the three-flavour condensate than the topological susceptibility for physical quark mass hierarchies, and thus offers the possibility to determine whether the three-flavour condensate is suppressed compared to the two-flavour one.
As an illustration, we provided numerical predictions for the two topological observables in the case of the results of the RBC and UKQCD collaborations already analysed in Ref.~\cite{Bernard:2012fw}. 

In addition, we discussed the case of three degenerate quarks, and we showed that a combination of the topological susceptibility and the fourth cumulant is able to pin down the thee-flavour quark condensate in a particularly clean way, as all next-to-leading order contributions cancel in this combination. We also considered the formulae for the two topological observables in two-flavour Chiral Perturbation Theory, and discussed the convention and physical content of the low-energy constants arising in the resulting chiral expansions.

Obviously, it remains to be seen if the fourth cumulant of the distribution of the topological charge can be computed accurately using lattice techniques. But we hope that our results will prompt lattice collaborations to study this topological observable in more detail, providing a new handle on the chiral structure of QCD vacuum.

\section{Acknowledgments}
Work supported in part by the European Community-Research  Infrastructure Integrating Activity "Study of Strongly Integrating Matter" (acronym
HadronPhysics3, Grant Agreement n. 283286) under the Seventh Framework Programme of EU.  

\appendix

\section{Diagram contributions to $c_4$}\label{app:oneloop}

As indicated in Sec.~\ref{sec:combinatorics}, there are several classes of diagrams contributing to the chiral expansion of $c_4$ up to one loop. In addition to the five LO diagrams in Fig.~\ref{figure:LOdiag} dressed by adding a counterterm or a loop at the level of the vertices or the propagators, there are five additional diagrams collected in Fig.~\ref{figure:NLOdiagother}.
In the following, we will give the contribution of the various diagrams with the same distinction for the argument of logarithms coming from tadpole or from unitarity contributions as in Eq.~(\ref{eq:c4NLO2logs}).

\vspace{0.2cm}

$\bullet$ Diagrams 1
\bea
&&-\frac{1}{81} B_0 F_0^2 (2 m+ms)+ \frac{ B_0^2 m^2}{216 \pi ^2} \lnmpikr
+\frac{ B_0^2
   \left(m+m_s\right){}^2}{648 \pi ^2} \lnmkkr
+\frac{ B_0^2 \left(m+2m_s\right){}^2}{5832 \pi ^2} \lnmetakr
 \\
&&
-\frac{128}{81} B_0^2  \left(m_s+2 m\right){}^2 L_6^r(\mu)
-\frac{128}{81}  B_0^2  \left(m_s+2
   m\right){}^2 L_7 -\frac{128}{81}  B_0^2  \left(2 m^2+m_s^2\right) L_8^r(\mu)
\nonumber\ena

$\bullet$ Diagrams 2
\bea
&&\frac{8}{81}B_0F_0^2 \frac{(m-m_s)^2}{(2m_s+m)}-\frac{ B_0^2  m^2 \left(m-m_s\right) \left(5 m_s+m\right)}{54 \pi ^2
   \left(m+2m_s\right){}^2}  \lnmpikr
\\
&&
   -
\frac{ B_0^2  \left(m-m_s\right)\left(m+m_s\right) \left(3m^2-8mm_s-7m_s^2\right){}^2}{324 \pi ^2
   \left(m+2m_s\right){}^2} \lnmkkr\nonumber\\
   &&
-\frac{ B_0^2  \left(m-m_s\right) \left(m^2-11 m m_s-8 m_s^2\right)}{1458 \pi ^2 \left(2
   m_s+m\right)} \lnmetakr
\nonumber \\
&&
+\frac{1024  B_0^2  \left(m_s+2 m\right) \left(m-m_s\right){}^2}{81 \left(m+2m_s\right)}
L_6^r(\mu)
+\frac{256 
   B_0^2  \left(8 m^2+29 m m_s+8 m_s^2\right) \left(m-m_s\right){}^2}{81 \left(m+2m_s\right){}^2}
L_7
\nonumber \\
&&
+\frac{256
    B_0^2  \left(4 m^2+15 m m_s+8 m_s^2\right) \left(m-m_s\right){}^2}{81 \left(m+2m_s\right){}^2} L_8^r(\mu)
\nonumber\ena

$\bullet$ Diagrams 3
\bea
&&
-\frac{4}{27}B_0^2 \frac{(m-m_s)^2}{27 (m+2m_s)}+\frac{ B_0^2  m^2 \left(m-m_s\right) \left(5 m_s+m\right)}{36 \pi ^2
   \left(m+2m_s\right){}^2} \lnmpikr
\\
&&
-\frac{ B_0^2  \left(m-m_s\right)\left(m+m_s\right) \left(7m^2-10mm_s-9m_s^2\right){}^2}{216 \pi^2 \left(m+2m_s\right){}^2} \lnmkkr
\nonumber\\&&
+\frac{ B_0^2  \left(m-m_s\right) \left(m^2-11 m m_s-8 m_s^2\right)}{972 \pi ^2 \left(2
   m_s+m\right)} \lnmetakr
\nonumber \\
&&
-\frac{256  B_0^2  \left(4 m^2+mm_s +4m_s^2 \right) \left(m-m_s\right){}^2}{27 \left(2
   m_s+m\right){}^2}L_6^r(\mu)
-\frac{512  B_0^2  \left(m_s+2 m\right) \left(m-m_s\right){}^2}{27 \left(2
   m_s+m\right)} L_7
\nonumber \\
&&
-\frac{256  B_0^2  \left(2 m^2+3 m m_s+4 m_s^2\right) \left(m-m_s\right){}^2}{27 \left(2
   m_s+m\right){}^2} L_8^r(\mu)
\nonumber\ena

$\bullet$ Diagrams 4
\bea
&&
\frac{8}{81}B_0F_0^2\frac{(m-4m_s)(m-m_s)^3}{(m+2m_s)^3}-\frac{ B_0^2  m^2
   \left(m^2+10 m m_s-38 m_s^2\right) \left(m-m_s\right){}^2}{54 \pi ^2 \left(m+2m_s\right){}^4} \lnmpikr  
\\
&&
-\frac{B_0^2  \left(m+m_s\right) \left(m-m_s\right)^2\left(18 m^3-121 m^2 m_s+113 m m_s^2+98
   m_s^3\right)} {648 \pi ^2 \left(m+2m_s\right){}^4} 
\lnmkkr  \nonumber\\
&&
-\frac{ B_0^2  \left(m^3-42 m^2 m_s+90 m m_s^2+32 m_s^3\right) \left(m-m_s\right){}^2}{1458 \pi ^2
   \left(m+2m_s\right){}^3} \lnmetakr
\nonumber\\
&&
+\frac{256  B_0^2  \left(8 m^2-mm_s-16m_s^2 \right) \left(m-m_s\right){}^3}{81 \left(2
   m_s+m\right){}^3} L_6^r(\mu)
 \nonumber\\
&&
+\frac{512  B_0^2  \left(4 m^3+21 m^2 m_s-63 m m_s^2-16 m_s^3\right)
   \left(m-m_s\right){}^3}{81 \left(m+2m_s\right){}^4}
L_7
\nonumber \\
&&
+\frac{256  B_0^2  \left(4 m^3+17 m^2m_s-52mm_s^2+32m_s^3\right) \left(m-m_s\right){}^3}{81 \left(m+2m_s\right){}^4} L_8^r(\mu)
\nonumber\ena

$\bullet$ Diagrams 5
\bea
&&
-\frac{2}{81}B_0F_0^2\frac{(m-m_s)^4(m+8m_s)}{(m+2m_s)^4}+
\frac{B_0^2  m^2 \left(m-m_s\right){}^3 
\left(m^2+13 m m_s+94 m_s^2\right)}{216 \pi ^2 \left(m+2m_s\right){}^5}
\lnmpikr
\\
&&
+\frac{B_0^2  \left(m-m_s\right){}^3 \left(m+m_s\right) \left(41m^3+133 m^2m_s-410 m m_s^2-304 m_s^3\right)}{3240 \pi
   ^2 \left(m+2m_s\right){}^5} \lnmkkr
\nonumber\\
&&
+\frac{B_0^2  \left(m-m_s\right){}^3 \left(m^3-3 m^2m_s-258 m m_s^2-64 m_s^3\right)}{5832 \pi ^2
   \left(m+2m_s\right){}^4} \lnmetakr
\nonumber\\
&&
+\frac{B_0^2  m^2 \left(m-m_s\right){}^4}{72 \pi ^2 
\left(m+2m_s\right){}^4} \lnmpi
+\frac{B_0^2  m^2 \left(m-m_s\right){}^4}
{216\pi ^2 \left(m+2m_s\right){}^4} \lnmk
\nonumber\\
&&
+\frac{B_0^2  \left(m-m_s\right){}^4 
\left(m+8m_s\right)){}^2}{1944 \pi ^2 \left(m+2m_s\right){}^4} \lnmeta
\nonumber\\
&&
-\frac{64  B_0^2  \left(8 m^2+41 m m_s+32 m_s^2\right) \left(m-m_s\right){}^4}{81 \left(2
   m_s+m\right){}^4} L_6^r(\mu)
\nonumber \\
&&
-\frac{256  B_0^2  \left(2 m^3+3 m^2 m_s+60 m m_s^2+16 m_s^3\right)
   \left(m-m_s\right){}^4}{81 \left(m+2m_s\right){}^5} L_7
\nonumber \\
&&
-\frac{256  B_0^2  \left(m^3+5 m^2m_s+32 mm_s^2+16m_s^3\right) \left(m-m_s\right){}^4}{81 \left(m+2m_s\right){}^5} L_8^r(\mu)
\nonumber\\
&&
+\frac{B_0^2 \left(m-m_s\right){}^4 \left(37 m^2+16 m m_s+64m_s^2\right)}{1944 \pi ^2 \left(m+2m_s\right){}^4}
\nonumber\ena

$\bullet$ Diagrams 6
\bea
&&
\frac{B_0^2}{72\pi^2}  m^2  \lnmpi
+\frac{B_0^2}{216 \pi^2} \left(m+m_s\right){}^2 \lnmk
+\frac{B_0^2}{1944 \pi^2}   \left(m+2m_s\right){}^2 \lnmeta
\\
&&
+\frac{B_0^2}{1944 \pi^2} \left(37 m^2+22 m m_s+13 m_s^2\right)
\nonumber\ena

$\bullet$ Diagrams 7
\bea
&&
-\frac{ B_0^2  m^2 \left(m-m_s\right)}{18 \pi^2 \left(m+2m_s\right)}  \lnmpi
-\frac{ B_0^2 m_s \left(m_s^2-m^2\right)}{54 \pi^2 \left(m+2m_s\right)} \lnmk
-\frac{B_0^2}{486 \pi^2}  \left(m-4 m_s\right) \left(m-m_s\right)  \lnmeta
\quad\\
&&
-\frac{ B_0^2 \left(m-m_s\right){}^2 \left(17 m_s+28 m\right)}{486 \pi^2 \left(m+2m_s\right)}
\nonumber\ena

$\bullet$ Diagrams 8
\bea
&&
\frac{ B_0^2  m^2
   \left(m-m_s\right){}^2}{18 \pi^2 \left(m+2m_s\right){}^2} \lnmpi
+\frac{ 
   B_0^2  m_s^2 \left(m-m_s\right){}^2}{54 \pi^2 \left(m+2m_s\right){}^2} \lnmk
+\frac{ B_0^2  \left(m-4 m_s\right){}^2 \left(m-m_s\right){}^2}{486 \pi^2 \left(m+2m_s\right){}^2}  \lnmeta
\quad \\
&&
+\frac{ B_0^2 \left(m-m_s\right){}^2 \left(28 m^2-8 m m_s+25 m_s^2\right)}
{486 \pi^2 \left(m+2m_s\right){}^2}
\nonumber\ena

$\bullet$ Diagrams 9
\bea
&&
\frac{ B_0^2
   m^2 \left(m-m_s\right){}^2}{36 \pi^2 \left(m+2m_s\right){}^2}  \lnmpi
-\frac{ B_0^2
  m\left(m-m_s\right)^2 \left(m+m_s\right) }{108  \pi^2 \left(m+2m_s\right){}^2} \lnmk
\\
&&
+\frac{ B_0^2 \left(m+8m_s\right)\left(m-m_s\right){}^2}{972  \pi^2 \left(m+2m_s\right)} \lnmeta
-\frac{ B_0^2
  \left(m-m_s\right)^2 \left(m+m_s\right)^2}{72 \pi^2 \left(m+2m_s\right){}^2} \lnmkkr
\nonumber\\
&&
+\frac{ B_0^2 \left(m-m_s\right){}^2 \left(19 m^2+mm_s+16m_s^2 \right)}{972 \pi^2 \left(m+2m_s\right){}^2}
\nonumber\ena

$\bullet$ Diagrams 10
\bea
&&
-\frac{
   B_0^2  m^2 \left(m-m_s\right){}^3}{18 \pi^2 \left(m+2m_s\right){}^3}  \lnmpi
-\frac{ B_0^2  m m_s \left(m-m_s\right){}^3}{54 \pi^2 \left(m+2m_s\right){}^3}  \lnmk
\\
&&
-\frac{ B_0^2  \left(m-4 m_s\right)\left(m+8m_s\right)
\left(m-m_s\right){}^3}{486 \pi^2 \left(2
   m_s+m\right){}^3}  \lnmeta
   -\frac{ B_0^2  (m +m_s) m_s \left(m-m_s\right){}^3}{36 \pi^2 \left(m+2m_s\right){}^3}  \lnmkkr
\nonumber\\
&&
-\frac{B_0^2 \left(m-m_s\right){}^3 \left(28 m^2+13 m m_s-32 m_s^2\right)}{486\pi^2 \left(m+2m_s\right){}^3}
\nonumber\ena

Let us notice that vertices from ${\mathcal L}_1$ with derivatives applied to the internal propagators generate contact terms which yield tadpole contributions even in the case of the scattering diagrams 9 and 10 where such contributions would not be expected naively. One can easily obtain the equivalent contributions with the prescription of Eq.~(\ref{eq:c4NLO}) by setting the LO masses $\mo_P^2$ in the argument of all logarithms. 

\section{Low-energy constants in Resummed Chiral Perturbation Theory} \label{app:L6L7L8}

In Resummed $\chi$PT~\cite{DescotesGenon:2003cg,DescotesGenon:2007ta,Bernard:2010ex,Bernard:2012fw}, the chiral expansions of $F_P^2$ and $F_P^2 M_P^2$ are expected to have small higher-order remainders when expressed in terms of the couplings arising in the chiral Lagrangian. These exact mass and decay constant identities can be inverted to reexpress NLO LECs in terms of LO parameters of the chiral Lagrangian, physical quantities, and HO remainders
\begin{eqnarray} 
\label{eq:deltal4}
Y(3)\Delta L_4 &=& \frac{1}{8(r+2)}\frac{F_\pi^2}{M_\pi^2}
  [1-\eta(r)-Z(3)-e]\,,
\\ 
\label{eq:deltal5}
Y(3)\Delta L_5 &=& \frac{1}{8}\frac{F_\pi^2}{M_\pi^2}
  [\eta(r)+e']\,, 
\\
\label{eq:deltal6}
Y^2(3)\Delta L_6 &=& \frac{1}{16(r+2)}\frac{F_\pi^2}{M_\pi^2}
  [1-\epsilon(r)-X(3)-d]\,,
\\ 
\label{eq:deltal8}
Y^2(3)\Delta L_8 &=& \frac{1}{16}\frac{F_\pi^2}{M_\pi^2}
  [\epsilon(r)+d']\,.
\end{eqnarray}
with
\begin{equation} \label{funcr}
\eta(r)=\frac{2}{r-1}\left(\frac{F_K^2}{F_\pi^2}-1\right)\,,
\qquad \epsilon(r) = 2\frac{r_2-r}{r^2-1}, \qquad
r_2= 2\left(\frac{F_KM_K}{F_{\pi}M_{\pi}}\right)^2 -1\sim 36\,,
\end{equation}
so that $\epsilon(r_2)=0$, and $\epsilon(r_1)=1$ with $r_1=2 (F_KM_K)/(F_\pi M_\pi)-1\simeq 8$.
$d$, $d'$, $e$ and $e'$ are combinations of  
HO remainders associated with the chiral expansions of $\pi$, $K$ masses and decay 
constants respectively.
$\Delta L_i=L_i^r(\mu)-\hat{L}_i(\mu)$ are independent of the renormalisation scale $\mu$
and combine the (renormalized and quark-mass independent)
constants $L_{4,5,6,8}$ together with chiral logarithms:
\begin{eqnarray}
32\pi^2\hat{L}_4(\mu)
          &=& \frac{1}{8}
          \log\frac{\mo_K^2}{\mu^2}
  -\frac{1}{8(r-1)(r+2)}
  \left[(4r+1)\log \frac{\mo_K^2}{\mo_\pi^2} 
      + (2r+1)\log \frac{\mo_\eta^2}{\mo_K^2}  \right]   
\label{eq:l4} \,,\\
32\pi^2 \hat{L}_5(\mu) &=& \frac{1}{8}
    \left[\log\frac{\mo_K^2}{\mu^2}+2\log\frac{\mo_\eta^2}{\mu^2}\right]
+\frac{1}{8(r-1)}
    \left[3\log\frac{\mo_\eta^2}{\mo_K^2}+5\log\frac{\mo_K^2}{\mo_\pi^2}\right]
\label{eq:l5} \,,\\
32\pi^2\hat{L}_6(\mu) &=& \frac{1}{16}\left[
       \log\frac{\mo_K^2}{\mu^2} 
       + \frac{2}{9}\log \frac{\mo_\eta^2}{\mu^2}
         \right] 
 -\frac{1}{16}\frac{r}{(r+2)(r-1)} \left[ 3 \log
          \frac{\mo_K^2}{\mo_\pi^2}  + \log \frac{\mo_\eta^2}{\mo_K^2} \right]
\label{eq:l6}\,, \\
32\pi^2 \hat{L}_8(\mu) &=& \frac{1}{16}
    \left[\log\frac{\mo_K^2}{\mu^2}+\frac{2}{3}\log\frac{\mo_\eta^2}{\mu^2}\right]
+\frac{1}{16(r-1)} \left[ 3 \log
          \frac{\mo_K^2}{\mo_\pi^2}  + \log \frac{\mo_\eta^2}{\mo_K^2} \right]
\label{eq:l8}\,.
\end{eqnarray}

Similarly, one can invert the relation for the $\eta$ mass, yielding
\begin{equation}\label{eq:L7}
[Y(3)]^2L_7 = \frac{1}{32(r-1)^2}\frac{F_\pi^2}{M_\pi^2}
 \Bigg[\frac{3F_\eta^2 M_\eta^2-4F_K^2M_K^2+F_\pi^2M_\pi^2}{F_\pi^2M_\pi^2}-d_{GO}-(r-1)^2[\epsilon(r)+d']
 \Bigg]\,,
\end{equation}
where $F_\eta$ is not measured, but can be computed using
\begin{eqnarray}\label{etadecay}
F_\eta^2 &=& F_\pi^2 Z(3)
  +8(r+2)Y(3)M_\pi^2 \Delta L_4
  +\frac{8}{3}(2r+1)Y(3)M_\pi^2 \Delta L_5\\
&&  +\frac{Y(3)M_\pi^2}{48\pi^2}\left[(2r+1)\log\frac{\mo_\eta^2}{\mo_K^2}-\log\frac{\mo_K^2}{\mo_\pi^2}\right]+
 F_\eta^2e_\eta\,, \nonumber
\end{eqnarray}
and we have introduced the difference of HO remainders
\begin{equation}
d_{GO}=\frac{3F_\eta^2M_\eta^2}{F_\pi^2M_\pi^2}d_\eta-
  \frac{4F_K^2M_K^2}{F_\pi^2M_\pi^2}d_K+d_\pi\,.
\end{equation}

\end{document}